\documentclass[final, 5p, times, twocolumn]{elsarticle}
\usepackage{amsmath,amsfonts,amssymb,bm}
\usepackage{color}
\usepackage{dsfont}
\usepackage{graphicx}
\usepackage{multirow}
\usepackage{soul}

\usepackage[mathscr,scaled=1.15]{urwchancal}
\DeclareFontFamily{OT1}{pzc}{}
\DeclareFontShape{OT1}{pzc}{m}{it}%
{<-> s * [1.15] pzcmi7t}{}
\DeclareMathAlphabet{\mathpzc}{OT1}{pzc}{m}{it}

\definecolor{purple}{rgb}{0.5,0,0.5}
\definecolor{blue}{rgb}{0.0,0,0.9}
\definecolor{prdblue}{rgb}{0.133,0.118,0.498}
\usepackage[colorlinks=true, pdfstartview=FitV, linkcolor=prdblue, citecolor= prdblue, urlcolor=prdblue]{hyperref}

\biboptions{sort&compress}

\journal{Physics Letters B}

\begin{document}

\begin{frontmatter}



\title{$\,$\\[-7ex]\hspace*{\fill}{\normalsize{\sf\emph{Preprint no}. NJU-INP 041/21}}\\[1ex]
Semileptonic $B_c \to \eta_c, J/\psi$ transitions}


\author[NJU,INP]{Zhao-Qian Yao}
%\email[]{cdroberts@nju.edu.cn}
\ead{zqyao@smail.nju.edu.cn}

\author[ECT]{Daniele Binosi}
\ead{binosi@ectstar.eu}

\author[NJU,INP]{Zhu-Fang Cui}
%\email[]{cdroberts@nju.edu.cn}
\ead{phycui@nju.edu.cn}

\author[NJU,INP]{Craig D. Roberts\corref{cor2}}
%\email[]{cdroberts@nju.edu.cn}
\ead{cdroberts@nju.edu.cn}
\cortext[cor2]{Corresponding Author}

\address[NJU]{
School of Physics, Nanjing University, Nanjing, Jiangsu 210093, China}
\address[INP]{
Institute for Nonperturbative Physics, Nanjing University, Nanjing, Jiangsu 210093, China}

\address[ECT]{
European Centre for Theoretical Studies in Nuclear Physics
and Related Areas, Villa Tambosi, Strada delle Tabarelle 286, I-38123 Villazzano (TN), Italy}

\begin{abstract}
Using a systematic, symmetry-preserving continuum approach to the Standard Model strong-interaction bound-state problem, we deliver parameter-free predictions for all semileptonic $B_c \to \eta_c, J/\psi$ transition form factors on the complete domains of empirically accessible momentum transfers.  Working with branching fractions calculated therefrom, the following values of the ratios for $\tau$ over $\mu$ final states are obtained:
$R_{\eta_c}=0.313(22)$ and $R_{J/\psi}=0.242(47)$.
Combined with other recent results, our analysis confirms a $2\sigma$ discrepancy between the Standard Model prediction for $R_{J/\psi}$ and the single available experimental result.
\\[1ex]
%% Date
\leftline{2021 April 18}
%\leftline{2021 April 01}
\end{abstract}


%%\item Research highlight 1
%%\item Research highlight 2


\end{frontmatter}


\section{Introduction}\label{SecIntro}
\unskip
The $B_c$ meson was discovered a little over twenty years ago \cite{Abe:1998fb}.  With mass $m_{B_c}= 6.2749(8)\,$GeV \cite{Zyla:2020zbs}, it lies below the threshold for $B D$ decay; and since it is an open flavour state, electromagnetic decays are forbidden.  Thus, within the Standard Model, only flavour-changing weak decays are possible.  Consequently, $B_c$ has a relatively long lifetime \cite{Zyla:2020zbs}:
\begin{equation}
\label{lifetime}
0.510(9)\,{\rm ps},
\end{equation}
%%  10^{-12}
which is, \emph{e.g}.\ ten-billion-times longer than that of the $\eta_c$.  These things make the $B_c$ an especially interesting system: it is the lightest open-flavour bound-state of the two heaviest quarks in Nature that are experimentally pliable; and lives long enough to make measurements possible.

Flavour-changing $B_c$ weak decays involve one of the following transitions: $\bar b \to \bar u$, $\bar b \to \bar c$, $c\to s$, $c\to d$.  Specific entries in the Cabbibo-Kobayashi-Maskawa (CKM) matrix modulate the strengths of these transitions.  Since $|V_{cs}|$ is the largest of the four that can be involved here, one may anticipate that $B_c\to B_s$ transitions dominate.  This expectation is supported by contemporary calculations, \emph{e.g}.\ Refs.\,\cite{Barik:2009zz, Zhang:2020dla, Xu:2021iwv}.
Another factor is the available phase space.  For instance, with $\eta_c$, $J/\psi$ final states, this is more than ten-times larger than for $B$; and such magnification may be sufficient to overwhelm the factor of roughly six suppression from $|V_{cb}|/|V_{cd}|$.  Calculations of the branching fractions ratio bear this out, \emph{e.g}.\ \cite{Xu:2021iwv}: ${\mathpzc B}_{B_c^+\to\eta_c \ell^+ \nu}/{\mathpzc B}_{B_c^+\to B^0 \ell^+ \nu_\ell} \approx 6$, where $\ell$ is a light lepton.  (This is a longstanding qualitative prediction \cite{Scora:1995ty, Gouz:2002kk}.)
Therefore, it is not surprising that the $B_c$ was discovered in decays to $J/\psi$ final states, especially given the narrow, prominent decay width for $J/\psi\to \ell^+\ell^-$.

\begin{figure}[t]
\includegraphics[width=0.42\textwidth]{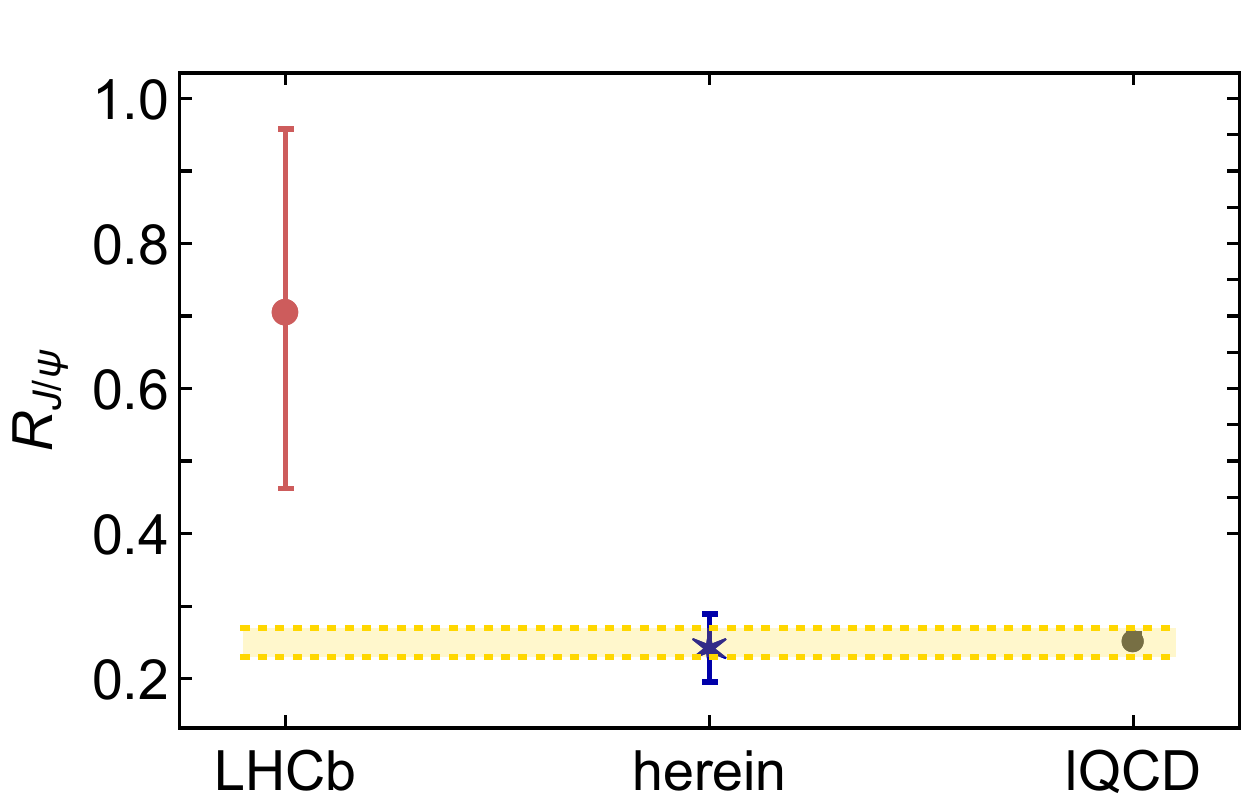}
\caption{\label{RJpsi}
Ratio $R_{J/\psi}$ in Eq.\,\eqref{eqRjpsi} -- red circle, empirical result from LHCb Collaboration \cite{Aaij:2017tyk};
blue star -- our prediction;
grey square -- lQCD result \cite{Harrison:2020nrv, Harrison:2020gvo};
and gold band -- unweighted mean of central values from contemporary calculations \cite{Tran:2018kuv, Issadykov:2018myx, Wang:2018duy, Leljak:2019eyw, Hu:2019qcn, Zhou:2019stx}
(Details provided below in connection with Table~\ref{TabBF}B.)
}
\end{figure}

Data acquired in the last decade, potentially indicating violations of lepton universality in $b$-quark decays
%
%%Measurement of an Excess of \bar{B} \to D^{(*)}\tau^- \bar{\nu}_\tau
%
\cite{Lees:2013uzd, Huschle:2015rga, Aaij:2015yra, Sato:2016svk, Hirose:2016wfn, Aaij:2017uff, Aaij:2021vac}, raise studies of the semileptonic decays of $B_c$-mesons with ground-state charmonia final-states to a new level of importance in the search for physics outside the Standard Model paradigm.  In fact, the LHCb collaboration has reported \cite{Aaij:2017tyk}:
\begin{equation}
\label{eqRjpsi}
R_{J/\psi} := \frac{{\mathpzc B}_{B_c^+\to J/\psi \tau \nu}}{{\mathpzc B}_{B_c^+\to J/\psi \mu \nu}} = 0.71 \pm 0.17 \,{\rm (stat)} \pm 0.18\,{\rm (syst)}
\end{equation}
and stated that this result lies approximately two standard-deviations ($2\sigma$) above the range of central values predicted by reliable calculations within the Standard Model, as highlighted by Fig.\,\ref{RJpsi}.  Such a discrepancy could signal violation of lepton universality in Nature's weak interactions.

Following early calculations \cite{Scora:1995ty}, numerous methods have been employed in the analysis of $B_c \to \eta_c, J/\psi$ semileptonic decays; and amongst the more recent are an array of continuum studies \cite{Tran:2018kuv, Issadykov:2018myx, Berns:2018vpl, Wang:2018duy, Leljak:2019eyw, Hu:2019qcn, Cohen:2019zev, Zhou:2019stx} and first results from lattice-regularised QCD (lQCD) \cite{Colquhoun:2016osw, Harrison:2020nrv, Harrison:2020gvo}.  We tackle the problem using a framework that is distinct from all these.  Namely, a continuum Schwinger function method (CSM) for solving hadron bound-state problems \cite{Horn:2016rip, Eichmann:2016yit, Fischer:2018sdj, Qin:2020rad}, which has provided a unified explanation for the properties of mesons and baryons with $0-3$ heavy quarks, \emph{i.e}.\ from Nature's (almost) Nambu-Goldstone bosons to triply-heavy baryons; see \emph{e.g}.\ Refs.\,\cite{Qin:2011dd, Qin:2011xq, Qin:2018dqp, Binosi:2014aea, Wang:2018kto, Ding:2018xwy, Binosi:2018rht, Qin:2019hgk, Xu:2019ilh, Yao:2020vef}.

\section{Transition Form Factors: Definitions}
\label{SecTFFs}
We consider the following transition matrix elements:
{\allowdisplaybreaks\begin{subequations}
\label{EqMEs}
\begin{align}
M_\mu^{B_c\to\eta_c}& (P,Q) =\langle \eta_c(p_{\eta_c}) | \bar c i\gamma_\mu b |B_c(k)\rangle \nonumber \\
& \hspace*{-1em}  = f_+(t) \, T_{\mu\nu}^Q P_\nu  + f_0(t)\, \tfrac{P\cdot Q}{Q^2} Q_\mu \,,\\
M_{\mu;\lambda}^{B_c\to J/\psi}&(P,Q) =\langle \psi^\lambda(p_{\psi};\lambda) | \bar c i(\gamma_\mu - \gamma_\mu \gamma_5) b |B_c(k)\rangle \nonumber \\
& \hspace*{-1em} =  2 m_{J/\psi} \tfrac{Q_\mu \epsilon^\lambda\cdot Q}{Q^2} A_0(t)
+  [m_{B_c}+m_{J/\psi}]T_{\mu\nu}^Q \epsilon_\nu^\lambda \, A_1(t)  \nonumber \\
&  + [P_\mu + Q_\mu \tfrac{m_{B_c}^2-m_{J/\psi}^2}{Q^2}]
\frac{\epsilon^\lambda\cdot Q \, A_2(t)}{m_{B_c}+m_{J/\psi}} \nonumber \\
& %+ Q_\mu \tfrac{\epsilon^\lambda\cdot Q \, A_3(t)}{m_{B_c}+m_{J/\psi}}
+ \varepsilon_{\mu\nu\rho\sigma}\epsilon_\nu^\lambda k_\rho p_{\psi \sigma}
\frac{2 V(t)}{m_{B_c}+m_{J/\psi}}\,,
\end{align}
\end{subequations}
where $Q^2T_{\mu\nu}^Q = Q^2\delta_{\mu\nu} - Q_\mu Q_\nu$,
$P =k+p_{\eta_c,\psi}$,
$Q=p_{\eta_c,\psi}-k$,
with $k^2 = -m_{B_{c}}^2$ and $p_{\eta_c}^2=-m_{\eta_c}^2$, $p_\psi^2=-m_{J/\psi}^2$;
$\epsilon_\nu^\lambda(p_f)$ is a polarisation four-vector, with $\sum_{\lambda=1}^3\epsilon_\nu^\lambda(p_f)\epsilon_\mu^\lambda(p_f) = T_{\mu\nu}^{p_f}$;
the squared-momentum-transfer is $t=-Q^2$;
and
$t_\pm^M = (m_{B_c} \pm m_M)^2$,
% =: m_{B_c}^2 y_-^{B_c} (M)$,
$M=\eta_c,J/\psi$.  ($t_-$ is the largest accessible value of $t$ in the identified physical decay process.)
}
The scalar functions in Eqs.\,\eqref{EqMEs} are the semileptonic transition form factors, which express all effects of hadron structure on the transitions.  %These are the quantities we calculate herein.
Ensuring the absence of kinematic singularities in Eqs.\,\eqref{EqMEs}, symmetries require
\begin{subequations}\label{EQsymmetries}
\begin{align}
f_+(0)& =f_0(0)\,,\\
A_0(0) & = \tfrac{m_{B_c} + m_{J/\psi}}{2m_{J/\psi}} A_1(0) - \tfrac{m_{B_c} - m_{J/\psi}}{2m_{J/\psi}} A_2(0)\,.
\end{align}
\end{subequations}

With predictions for the transition form factors in hand, one can compute the associated decay branching fractions from the differential decay width for $B_c \to M {\mathpzc l}^+ \nu_{\mathpzc l}$:
\begin{align}
\label{dGdt}
\left.\frac{d\Gamma}{dt}\right|_{B_c\to M{\mathpzc l}\nu_{\mathpzc l}} &
= \frac{G_F^2 |V_{cb}|^2}{192\pi^3 m_{B_c}^3}
\lambda(m_{B_c},m_M,t) \frac{(t-m_{\mathpzc l}^2)^2}{t^2}\, {\mathpzc H}^2,
\end{align}
where:
%$M=\eta_c, J/\psi$;
$G_F = 1.166 \times 10^{-5}\,$GeV$^{-2}$;
%\begin{equation}
$|V_{cb}| = 0.0410(14)$ \mbox{\cite{Zyla:2020zbs}};
%41:0(1:4)  10􀀀3
%\end{equation}
$\lambda(m_{B_c},m_M,t)^2 = (t_+ - t)(t_- - t)$;
\begin{align}
{\mathpzc H}^2 & = (H_+^2 + H_-^2 + H_0^2)(1+\tfrac{m_{\mathpzc l}^2}{2t})
+ \tfrac{3 m_{\mathpzc l}^2}{2 t} H_t^2,
\end{align}
$m_{\mathpzc l}^2\leq t \leq t_-$, $m_{\mathpzc l}$ is the lepton mass.
For $M=\eta_c$, $H_\pm \equiv 0$,
\begin{equation}
H_0 = \lambda(m_{B_c},m_{\eta_c},t) f_+(t)\,,\;
H_t = (m_{B_c}^2 - m_{\eta_c}^2) f_0(t)\,;
\end{equation}
whereas when $M=J/\psi$,
\begin{subequations}
\label{Hpsi}
\begin{align}
\tfrac{1}{\surd t}H_\pm & = (m_{B_c}+m_{J/\psi})A_1 (t) \mp \frac{\lambda(m_{B_c},m_{J/\psi},t)}{m_{B_c}+m_{J/\psi}} V(t) \,,\\
H_0 & = \frac{1}{2 m_{J/\psi}} \left[
(m_{B_c}^2-m_{J/\psi}^2 -t)(m_{B_c}+m_{J/\psi}) A_1(t) \right.   \nonumber \\
& \quad \left. - \frac{\lambda(m_{B_c},m_{J/\psi},t)^2}{m_{B_c}+m_{J/\psi}} A_2(t)\right],\\
H_t & =\lambda(m_{B_c},m_{J/\psi},t)\, A_0(t)\,.
\end{align}
\end{subequations}
After integrating Eq.\,\eqref{dGdt} to obtain the required partial widths, one quotes the branching fractions, ${\mathpzc B}_{B_c\to M{\mathpzc l} \nu_{\mathpzc l}}$, with respect to the total width determined from Eq.\,\eqref{lifetime}.

\section{Transition Form Factors: Matrix Elements}
\label{SecRL}
At leading-order (rainbow-ladder, RL) in the most widely used CSM truncation \cite{Munczek:1994zz, Bender:1996bb}, which has been employed to unify, \emph{inter alia}, all semileptonic pseudoscalar-to-pseudoscalar transitions involving $\pi$, $K$, $D_{(s)}$ initial states \cite{Ji:2001pj, Yao:2020vef}, the matrix elements in Eqs.\,\eqref{EqMEs} take the following form:
\begin{align}
\nonumber
M_\mu^{B_c\to M}&(P,Q)  = N_c {\rm tr}\int\frac{d^4 {\mathpzc s}}{(2\pi)^4}
S_b({\mathpzc s}-k) \Gamma_{B_c}({\mathpzc s}-k/2;k) S_c({\mathpzc s}) \\
& \times \Gamma_M({\mathpzc s}-p/2;-p) S_c({\mathpzc s}-p)
i {\mathpzc W}_\mu^{cb}({\mathpzc s}-p,{\mathpzc s}-k) \,,
\label{dMDs}
\end{align}
where $N_c=3$ and the trace is over spinor indices.
There are three types of matrix-valued functions in Eq.\,\eqref{dMDs}.  The simplest are the propagators for the dressed-quarks involved in the transition process: $S_f({\mathpzc s})$, $f=c,b$; then there are the Bethe-Salpeter amplitudes for the mesons involved: $\Gamma_{M}$; and, finally, the dressed $b\to c$ weak transition vertex: ${\mathpzc W}_\mu^{bc}$.  Each of these quantities can be computed once the kernel of the RL Bethe-Salpeter equation is specified.
Importantly, Eq.\,\eqref{dMDs} preserves the identities in Eqs.\,\eqref{EQsymmetries}, both algebraically and numerically.

With a realistic kernel, RL truncation provides a sound description of systems wherein \cite{Qin:2020rad}: (\emph{i}) orbital angular momentum does not play a large role and (\emph{ii}) the non-Abelian anomaly can be neglected.  In such cases, corrections to the truncation largely cancel amongst themselves.  $\eta_c$, $J/\psi$, $B_c$ are amongst the systems for which these conditions hold.  Herein, we use the RL kernel detailed in Refs.\,\cite{Qin:2011dd, Qin:2011xq, Qin:2018dqp}:
 \begin{subequations}
\label{KDinteraction}
\begin{align}
\mathscr{K}_{\rho_1\rho_1',\rho_2\rho_2'} & = {\mathpzc G}_{\mu\nu}(k) [i\gamma_\mu]_{\rho_1\rho_1'} [i\gamma_\nu]_{\rho_2\rho_2'}\,,\\
 {\mathpzc G}_{\mu\nu}(k) & = \tilde{\mathpzc G}(k^2) T_{\mu\nu}^k\,,
\end{align}
\end{subequations}
with ($s=k^2$)
\begin{align}
\label{defcalG}
 %\tfrac{1}{Z_2^2} = 1 when zeta=zeta_19
 \tilde{\mathpzc G}(s) & =
 \frac{8\pi^2}{\omega^4} D e^{-s/\omega^2} + \frac{8\pi^2 \gamma_m \mathcal{F}(s)}{\ln\big[ \tau+(1+s/\Lambda_{\rm QCD}^2)^2 \big]}\,,
\end{align}
%KDinteractiondefcalG
where $\gamma_m=4/\beta_0$, $\beta_0=11-(2/3)n_f$, $n_f=5$, $\Lambda_{\rm QCD} = 0.36\,$GeV, $\tau={\rm e}^2-1$, and ${\cal F}(s) = \{1 - \exp(-s/[4 m_t^2])\}/s$, $m_t=0.5\,$GeV.  %$Z_2$ is the computed dressed-quark wave function renormalisation constant.
Following standard practice, in solving all integral equations \cite{Chang:2008ec}, we use a mass-independent momentum-subtraction renormalisation scheme, fixing each renormalisation constant in the chiral limit, with renormalisation scale $\zeta=19\,$GeV$=:\zeta_{19}$.

The elaboration of Eqs.\,\eqref{KDinteraction}, \eqref{defcalG} and their connection with QCD are described in Refs.\,\cite{Qin:2011dd, Qin:2011xq, Binosi:2014aea}.  Here, we simply reiterate some points.
(\emph{i}) The interaction is consistent with that found through studies of QCD's gauge sector, capitalising on the fact that the gluon propagator is a bounded, smooth function of spacelike momenta, which achieves its maximum value on this domain at $s=0$ \cite{Binosi:2016xxu, Gao:2017uox, Cui:2019dwv}; and the dressed gluon-quark vertex does not possess any structure which can qualitatively alter these features \cite{Kizilersu:2021jen}.
(\emph{ii}) It is specified in Landau gauge because, \emph{inter alia}, this gauge is a fixed point of the renormalisation group and ensures that sensitivity to the form of the gluon-quark vertex is minimal, thus providing the conditions for which RL truncation is most accurate.  %One may subsequently appeal to gauge covariance of $n$-point Schwinger functions to ensure that results are independent of the gauge choice.
(\emph{iii}) The interaction preserves the one-loop renormalisation group behaviour of QCD; hence, \emph{e.g}.\ the quark mass-functions produced are independent of the renormalisation point.
(\emph{iv}) On $s \lesssim (2 m_t)^2$, Eq.\,\eqref{defcalG} defines a two-parameter \emph{Ansatz}, the details of which determine whether such corollaries of emergent hadron mass as confinement and dynamical chiral symmetry breaking are realised in solutions of the bound-state equations \cite{Roberts:2020hiw, Roberts:2021nhw}.

The analyses in Ref.\,\cite{Qin:2018dqp} determined that one can unify the properties of a diverse range of systems using $\omega = 0.8\,$GeV, $\varsigma^3=D\omega = (0.6\,{\rm GeV})^3$ and we use these values hereafter.  An additional feature of Eq.\,\eqref{defcalG} is that with a given value of $\varsigma$, results for observable quantities are practically insensitive to variations $\omega \to \omega (1\pm 0.1)$; so, there is no issue of fine tuning.

\begin{table}[t]
\caption{\label{psmesonstatic}
Static properties of mesons evaluated using bound-state equations defined by the kernel specified in Eqs.\,\eqref{KDinteraction}, \eqref{defcalG}.
Normalisation: the empirical value of the pion's leptonic decay constant is $f_\pi \approx 0.092\,$GeV.
Empirical values (expt.), where available, drawn from Ref.\,\cite{Zyla:2020zbs}; and lattice-QCD (lQCD) results for leptonic decay constants from Refs.\,\cite{Davies:2010ip, McNeile:2012qf, Colquhoun:2015oha}.
The mean absolute relative error between our predictions and empirical results is 3.6\%.
(All dimensioned quantities in GeV.)}
\begin{center}
\begin{tabular*}%{|c|c|c|c|c|c|c|}\hline
{\hsize}
{
l@{\extracolsep{0ptplus1fil}}
|c@{\extracolsep{0ptplus1fil}}
c@{\extracolsep{0ptplus1fil}}
c@{\extracolsep{0ptplus1fil}}
c@{\extracolsep{0ptplus1fil}}
c@{\extracolsep{0ptplus1fil}}
c@{\extracolsep{0ptplus1fil}}
c@{\extracolsep{0ptplus1fil}}
c@{\extracolsep{0ptplus1fil}}
c@{\extracolsep{0ptplus1fil}}
c@{\extracolsep{0ptplus1fil}}}\hline
 & $m_{\eta_c}$ & $m_{J/\psi}$ & $m_{B_c}$ & $m_{\eta_b}$ & $m_{\Upsilon}$
 & $f_{\eta_c}$ & $f_{J/\psi}$ & $f_{B_c}$ & $f_{\eta_b}$ & $f_{\Upsilon}$ \\ \hline
herein$\ $&$2.98\ $ & $3.12\ $ & $6.27\ $ & $9.19\ $ & $9.28\ $ &
$0.28\ $ & $0.30\ $ & $0.43\ $ & $0.56\ $ & $0.53\ $ \\\hline
expt.$\ $ &$2.98\ $ & $3.10\ $ & $6.27\ $ & $9.40\ $ & $9.46\ $ &
$0.24$ & $0.29\ $ & &  & $0.51$ \\\hline
lQCD$\ $ & & & & & &
$0.28$ & $0.29\ $ & $0.31\ $ & $0.47\ $& $0.46\ $ \\\hline
\end{tabular*}
\end{center}
\end{table}
%% Davies:2010ip
%% McNeile:2012qf
%% Colquhoun:2014ica

We now illustrate the qualities of the framework by computing an array of heavy pseudoscalar meson static properties, \emph{i.e}.\ their masses and leptonic decay constants.  Solving the gap and Bethe-Salpeter equations (see, \emph{e.g}.\ Ref.\,\cite{Xu:2019ilh} and Ref.\,\cite[Appendix\,1]{Yao:2020vef}) with the following values of the renormalisation-point-invariant $c$ and $b$ current-quark masses (in GeV):
\begin{equation}
\label{hatmcb}
\hat m_c = 1.61\,,\;
\hat m_b = 6.96\,,
\end{equation}
one obtains the results in Table~\ref{psmesonstatic}.  The mean absolute relative error between our predictions and empirical values is 3.6\%.  For later use, we note that $r_{b:c}=\hat m_{b}/\hat m_{c}=4.32$ and, equivalently, $r_{c:b}=0.23$.

The masses in Eq.\,\eqref{hatmcb} correspond to the following current masses at our renormalisation scale:
$m_c^{\zeta_{19}}=0.82\,$GeV, $m_b^{\zeta_{19}}=3.55\,$GeV;
and one-loop evolved to $\zeta=\zeta_2=2\,$GeV,
$m_c^{\zeta_{2}}=1.22\,$GeV, $m_b^{\zeta_{2}}=5.26\,$GeV.  Working with the dressed-quark mass-functions obtained by solving the gap equations, $M_{c,b}(k)$, and defining Euclidean constituent-quark masses as the solutions of $M_{c,b}(M^E_{c,b}) = M^E_{c,b}$, one finds (in GeV):
\begin{equation}
M^E_c = 1.33\,,\;
M^E_b = 4.12\,.
\end{equation}
These quantities are analogous to the ``running masses'' often quoted in connection with heavy quarks and our predictions are within 3\% of those listed elsewhere \cite{Zyla:2020zbs}.

It is worth remarking on some important physical aspects of the weak transition vertex, ${\mathpzc W}_\mu^{cb}$.  Pseudoscalar-to-pseudoscalar transitions only involve the vector part, which possesses poles at $Q^2 + m_{B_c^\ast,B_{c0}^\ast}^2=0$.  Pseudoscalar-to-vector transitions also involve the axial-vector part.  This has poles at $Q^2 + m_{B_c,  B_{c1}}^2=0$.  The presence of these poles is a prerequisite for any valid analysis of $B_c \to \eta_c, J/\psi$ semileptonic transitions.  They are manifest in our treatment.

\section{Computational Scheme and Results}
The integration in Eq.\,\eqref{dMDs} samples the appearing functions on a material domain of their complex-valued arguments.  So long as the masses of the initial and final state mesons are similar, \emph{i.e}.\ the ratio of the current-masses of the quarks involved, $r_{Q_1:Q_2}$, does not differ too much from unity, the integral can readily be evaluated using simple numerical techniques because $t_-$ and, hence, the maximum momentum of the recoiling meson, remains modest.  However, at some value of $r_{Q_1:Q_2}=:r_f$, $t_-$ becomes so large that singularities associated with the analytic structure of the dressed-quark propagators \cite{Maris:1997tm, Windisch:2016iud} move into the complex-${\mathpzc s}^2$ integration domain and straightforward numerical techniques fail.

%Solutions for meson Bethe-Salpeter amplitudes and inhomogeneous vertices are readily obtained when the ratio of the current-masses of the quarks involved, $r_{Q_1:Q_2}$, does not differ too much from unity.  However, at some value of $r_{Q_1:Q_2}=:r_f$, owing to the analytic structure of the dressed-quark propagators and related moving singularities in the complex-${\mathpzc s}^2$ domain sampled by the bound-state equations \cite{Maris:1997tm, Windisch:2016iud}, simple methods fail.

This sort of problem was solved in Ref.\,\cite{Chang:2013nia} by using perturbation theory integral representations (PTIRs) \cite{Nakanishi:1969ph} for each matrix-valued function in the integrand defining the associated matrix element.  However, constructing accurate PTIRs is time consuming;
%%%
% P->P ... S*3 = 6 + P*2 = 8 + GcbV = 12
% P->V ... new ... V = 8 + GcbA = 12
%%%
and here the challenge is compounded because the complete set of integrands involves 46 distinct scalar functions.  Like Ref.\,\cite{Yao:2020vef}, we therefore adopt a different approach.

(\emph{I}) -- We consider the semileptonic transitions of a fictitious $c \bar{\mathpzc Q}$ pseudoscalar meson: $B_{c \bar {\mathpzc Q}}\to \eta_{c\bar c}, J/\psi_{c\bar c}$.  All relevant Schwinger functions and, subsequently, the transition form factors are computed as a function of $\hat m_{\mathpzc Q}$ as it is increased from the point $r_{{\mathpzc Q}:c}=1$ to $r_f^{\eta_c}=3.17$ or $r_f^{J/\psi}=2.93$.
Then, using the statistical Schlessinger point method (SPM), exploited successfully elsewhere \cite{Chen:2018nsg, Binosi:2018rht, Binosi:2019ecz, Ding:2019lwe, Ding:2019qlr, Souza:2019ylx, Xu:2019ilh, Cui:2020rmu, Huber:2020ngt, Cui:2021vgm}, we build $\hat m_{\mathpzc Q}$-interpolations of all transition form factors, which are then used to extrapolate every measurable feature of the matrix elements to the physical point $r_{b:c}=4.32$, Eq.\,\eqref{hatmcb}.

%(\emph{I}) -- We consider the semileptonic transitions of a fictitious ${\mathpzc Q} \bar b$ pseudoscalar meson, $B_{{\mathpzc Q}\bar b}\to \eta_{{\mathpzc Q}\bar {\mathpzc Q}}, J/\psi_{{\mathpzc Q}\bar {\mathpzc Q}}$, where the final states are ${\mathpzc Q}\bar {\mathpzc Q}$ mesons, with $\eta_c, J/\psi$ quantum numbers.  All relevant Schwinger functions and, subsequently, the transition form factors are computed as a function of $\hat m_{\mathpzc Q}$ as it is reduced from the point $r_{{\mathpzc Q}:b}=1$ to $r_f^{\eta_{{\mathpzc Q}\bar {\mathpzc Q}}}=0.56$ or $r_f^{J/\psi_{{\mathpzc Q}\bar {\mathpzc Q}}}=0.63$.
%
%Then, using the statistical Schlessinger point method (SPM), exploited successfully elsewhere \cite{Chen:2018nsg, Binosi:2018rht, Binosi:2019ecz, Ding:2019lwe, Ding:2019qlr, Souza:2019ylx, Xu:2019ilh, Cui:2020rmu, Huber:2020ngt, Cui:2021vgm}, we build $\hat m_{\mathpzc Q}$-interpolations of all transition form factors, which are then used to extrapolate every measurable feature of the matrix elements to the physical point $r_{c:b}=0.23$,  Eq.\,\eqref{hatmcb}.

(\emph{II}) -- These exercises are repeated from an inverted perspective.  To wit, beginning with an analogous initial state, we consider $B_{{\mathpzc Q}\bar b}\to \eta_{{\mathpzc Q}\bar {\mathpzc Q}}, J/\psi_{{\mathpzc Q}\bar {\mathpzc Q}}$, where the final states are ${\mathpzc Q}\bar {\mathpzc Q}$ mesons, with $\eta_c, J/\psi$ quantum numbers.  The transition form factors are then computed as a function of $\hat m_{\mathpzc Q}$, reducing it from the point $r_{{\mathpzc Q}:b}=1$ to $r_f^{\eta_{{\mathpzc Q}\bar {\mathpzc Q}}}=0.56$ or $r_f^{J/\psi_{{\mathpzc Q}\bar {\mathpzc Q}}}=0.63$.  SPM extrapolation is subsequently used to reach the physical value, $r_{c:b}=0.23$,  Eq.\,\eqref{hatmcb}.  Since extrapolations of initial and final states are required here, the SPM uncertainty is larger.

%$B_{c\bar {\mathpzc Q}}\to \eta_{c\bar c}, J/\psi_{c\bar c}$ and working up from $r_{{\mathpzc Q}:c}=1$ to $r_f^{\eta_c}=3.17$ or $r_f^{J/\psi}=2.93$.  The physical value of $r_{b:c}=4.32$, Eq.\,\eqref{hatmcb}, is then reached by SPM extrapolation.

(\emph{III}) -- Having completed these exercises, we combine the outcomes to produce our final results.

It is worth remarking that the SPM is founded on interpolation via continued fractions \cite{Schlessinger:1966zz, PhysRev.167.1411}.  It is typically augmented today by statistical sampling.   The approach avoids any assumptions on the function used for the representation of input and captures both local and global features of that source.  This latter aspect underpins the reliability of subsequent extrapolations.  The SPM can accurately reconstitute a complex-valued function within a radius of convergence determined by that one of the function's branch points which lies nearest the real domain from which the sample points are drawn.  The statistical aspect ensures that one has a genuine estimate of the uncertainty associated with any extrapolation.

\begin{table}[t]
\caption{\label{SPMparameters}
\emph{Panels} {\sf A}-{\sf F}. SPM interpolation parameters for each transition form factor considered herein, as labelled: Eq.\,\eqref{TFfunction}, $\alpha_1$ is dimensionless and $\alpha_{2,3}$ have dimension GeV$^{-1}$.  \emph{N.B}.\ Regarding $f_{+,0}^{B_c\to \eta_c}$, $\alpha_1$ is the same in both cases because $f_+(0)=f_0(0)$.
(The SPM uncertainty estimate is discussed in the paragraph before that containing Eq.\,\eqref{mBcSPM}.)
}
\begin{center}
\begin{tabular*}%{|c|c|c|c|c|c|c|}\hline
{\hsize}
{
l@{\extracolsep{0ptplus1fil}}
|c@{\extracolsep{0ptplus1fil}}
c@{\extracolsep{0ptplus1fil}}
c@{\extracolsep{0ptplus1fil}}}\hline
({\sf A}) $f_+^{B_c\to \eta_c}\ $  & $\alpha_1\ $ & $\alpha_2\ $ & $\alpha_3\ $ \\ \hline
SPM (\emph{I})$\ $ & $0.634(07)\ $ & $0.0327(05)\ $ & $0.0550(08)\ $\\
SPM (\emph{II})$\ $ & $0.630(17)\ $ & $0.0318(33)\ $ & $0.0659(16)\ $ \\\hline
mean (\emph{III})$\ $ & $0.632(13)\ $ & $0.0323(24) \ $ & $0.0605(13)\ $\\ \hline
\end{tabular*}
\smallskip

\begin{tabular*}%{|c|c|c|c|c|c|c|}\hline
{\hsize}
{
l@{\extracolsep{0ptplus1fil}}
|c@{\extracolsep{0ptplus1fil}}
c@{\extracolsep{0ptplus1fil}}
c@{\extracolsep{0ptplus1fil}}}\hline
({\sf B}) $f_0^{B_c\to \eta_c}\ $  & $\alpha_1\ $ & $\alpha_2\ $ & $\alpha_3\ $ \\ \hline
SPM (\emph{I})$\ $ & $0.634(07)\ $ & $0.0243(08)\ $ & $0.0328(7)\ $\\
SPM (\emph{II})$\ $ & $0.630(17)\ $ & $0.0258(14)\ $ & $0.0352(6)\ $ \\\hline
mean (\emph{III})$\ $ & $0.632(13)\ $ & $0.0251(11) \ $ & $0.0340(7)\ $\\ \hline
\end{tabular*}
\smallskip

\begin{tabular*}%{|c|c|c|c|c|c|c|}\hline
{\hsize}
{
l@{\extracolsep{0ptplus1fil}}
|c@{\extracolsep{0ptplus1fil}}
c@{\extracolsep{0ptplus1fil}}
c@{\extracolsep{0ptplus1fil}}}\hline
({\sf C}) $A_0^{B_c\to J/\psi}\ $  & $\alpha_1\ $ & $\alpha_2\ $ & $\alpha_3\ $ \\ \hline
SPM (\emph{I})$\ $ & $0.563(13)\ $ & $0.0331(21)\ $ & $0.0572(049)\ $\\
SPM (\emph{II})$\ $ & $0.586(44)\ $ & $0.0269(58)\ $ & $0.0531(110)\ $ \\\hline
mean (\emph{III})$\ $ & $0.574(33)\ $ & $0.0300(43) \ $ & $0.0552(085)\ $\\ \hline
\end{tabular*}
\smallskip

\begin{tabular*}%{|c|c|c|c|c|c|c|}\hline
{\hsize}
{
l@{\extracolsep{0ptplus1fil}}
|c@{\extracolsep{0ptplus1fil}}
c@{\extracolsep{0ptplus1fil}}
c@{\extracolsep{0ptplus1fil}}}\hline
({\sf D}) $A_1^{B_c\to J/\psi}\ $  & $\alpha_1\ $ & $\alpha_2\ $ & $\alpha_3\ $ \\ \hline
SPM (\emph{I})$\ $ & $0.546(19)\ $ & $0.0175(29)\ $ & $0.0238(54)\ $\\
SPM (\emph{II})$\ $ & $0.557(29)\ $ & $0.0199(45)\ $ & $0.0488(68)\ $ \\\hline
mean (\emph{III})$\ $ & $0.551(25)\ $ & $0.0187(38) \ $ & $0.0363(61)\ $\\ \hline
\end{tabular*}
\smallskip

\begin{tabular*}%{|c|c|c|c|c|c|c|}\hline
{\hsize}
{
l@{\extracolsep{0ptplus1fil}}
|c@{\extracolsep{0ptplus1fil}}
c@{\extracolsep{0ptplus1fil}}
c@{\extracolsep{0ptplus1fil}}}\hline
({\sf E}) $A_2^{B_c\to J/\psi}\ $  & $\alpha_1\ $ & $\alpha_2\ $ & $\alpha_3\ $ \\ \hline
SPM (\emph{I})$\ $ & $0.546(11)\ $ & $0.0197(43)\ $ & $0.0508(68)\ $\\
SPM (\emph{II})$\ $ & $0.576(38)\ $ & $0.0213(24)\ $ & $0.0238(86)\ $ \\\hline
mean (\emph{III})$\ $ & $0.561(28)\ $ & $0.0205(35) \ $ & $0.0373(77)\ $\\ \hline
\end{tabular*}
\smallskip

\begin{tabular*}%{|c|c|c|c|c|c|c|}\hline
{\hsize}
{
l@{\extracolsep{0ptplus1fil}}
|c@{\extracolsep{0ptplus1fil}}
c@{\extracolsep{0ptplus1fil}}
c@{\extracolsep{0ptplus1fil}}}\hline
({\sf F}) $V^{B_c\to J/\psi}\ $  & $\alpha_1\ $ & $\alpha_2\ $ & $\alpha_3\ $ \\ \hline
SPM (\emph{I})$\ $ & $0.827(26)\ $ & $0.0439(54)\ $ & $0.0350(135)\ $\\
SPM (\emph{II})$\ $ & $0.845(51)\ $ & $0.0463(99)\ $ & $0.0210(088)\ $ \\\hline
mean (\emph{III})$\ $ & $0.836(41)\ $ & $0.0451(80) \ $ & $0.0280(110)\ $\\ \hline
\end{tabular*}
\end{center}
\end{table}

To elucidate further, we first compute the value of a given quantity, ${\mathpzc X}$, at $N=40$ different values of the evolving mass, $\hat m_{\mathpzc Q}$, distributed evenly on the domain of direct computation.
Then, $M=20$ values of $\hat m_{\mathpzc Q}$ are chosen at random from this $40$-element set, using which a continued fraction interpolation is developed for ${\mathpzc X}(\hat m_{\mathpzc Q})$ on this $20$-element subset.
A large number of interpolating functions, $n_I$, is subsequently obtained by inspecting the $C(N,M)$ combinatorial possibilities for the $M$ element subset and eliminating those functions which fail to satisfy a simple physical constraint; namely, we insist that each interpolation be smooth on the domain of required current-quark masses.  For all quantities considered, this constraint yields $n_I\approx 100,000$ acceptable interpolations.
Our prediction for ${\mathpzc X}$ is then obtained by extrapolating each of the associated $n_I$ physical SPM interpolants to the target current-mass and reporting as the result that value which sits at the centre of the band within which 68\% of the interpolants lie.  This $1\sigma$ band is quoted as the uncertainty in the result.

\begin{table}[t]
\caption{\label{SPMparametersMasses}
SPM predictions for meson masses (in GeV) that determine the locations of the timelike pole in the transition form factors computed herein, Eq.\,\eqref{TFfunction}.  These masses have not yet been measured; so, we present lQCD results for context \cite{Mathur:2018epb}.
(The SPM uncertainty estimate is discussed in the paragraph before that containing Eq.\,\eqref{mBcSPM}.)
}
\begin{center}
\begin{tabular*}%{|c|c|c|c|c|c|c|}\hline
{\hsize}
{
l@{\extracolsep{0ptplus1fil}}
|c@{\extracolsep{0ptplus1fil}}
c@{\extracolsep{0ptplus1fil}}
c@{\extracolsep{0ptplus1fil}}}\hline
 & $m_{B_c^\ast}\ $ & $m_{B_{c0}^\ast}\ $ & $m_{B_{c1}}\ $ \\ \hline
SPM (\emph{I})$\ $ & $6.402(20)\ $ & $6.767(21)\ $ & $6.880(20)\ $\\
SPM (\emph{II})$\ $ & $6.382(26)\ $ & $6.752(28)\ $ & $6.851(26)\ $ \\\hline
mean (\emph{III})$\ $ & $6.392(23)\ $ & $6.760(25) \ $ & $6.866(23)\ $\\ \hline
lQCD  & $6.331(07)\ $& $ 6.712(19) \ $& $6.736(18)\ $ \\\hline
\end{tabular*}
\end{center}
\end{table}

The reliability of our SPM procedure is readily illustrated.  The meson masses in Table~\ref{psmesonstatic} were computed directly via the Bethe-Salpeter equation using the masses in Eq.\,\eqref{hatmcb}.  One may equally compute masses using the procedures described in (\emph{I}) and (\emph{II}) above.
Using (\emph{I}) on $r_f^{\eta_{{\mathpzc Q}\bar {\mathpzc Q}}} \leq r_{{\mathpzc Q}:b} \leq 1$, we find $m_{B_c} = 6.259(1)\,$GeV; and employing (\emph{II}) on $1 \leq r_{{\mathpzc Q}:c} \leq r_f^{\eta_c}$, $m_{B_c} = 6.281(6)\,$GeV.  Hence, the final SPM result is
\begin{equation}
\label{mBcSPM}
m_{B_c} = 6.270(4)\,{\rm GeV},
\end{equation}
matching the directly computed value in Table~\ref{psmesonstatic}.
Repeating this exercise using the limiting current-masses in the $J/\psi$ channel, the SPM result is $6.267(8)\,$GeV, again agreeing with Table~\ref{psmesonstatic}.

On the physical domain associated with any value of $r_{Q_1:Q_2}$, each of the transition form factors can accurately be interpolated using the following function:
\begin{equation}
\label{TFfunction}
{\mathpzc f}(t) = \alpha_1 +  \alpha_2 \, t+ \frac{\alpha_3\,  t^2}{{\mathpzc m}^2-t}\,,
\end{equation}
where $\alpha_{1,2,3}$ and ${\mathpzc m}$ are functions of $r_{{\mathpzc Q}:b}$ or $r_{{\mathpzc Q}:c}$.  It is the coefficients $\alpha_{1,2,3}$ for which we develop SPM interpolations.  The results are listed in Table~\ref{SPMparameters}.

As noted in closing Sec.\,\ref{SecTFFs}, the pole masses in Eq.\,\eqref{TFfunction} correspond to particular mesons:
${\mathpzc f}=A_0$, then ${\mathpzc m}=m_{B_c}$;
${\mathpzc f}=f_+, V$, then ${\mathpzc m}=m_{B_c^\ast}$;
${\mathpzc f}=f_0$, then ${\mathpzc m}=m_{B_{c0}^\ast}$;
and
${\mathpzc f}=A_{1,2}$, then ${\mathpzc m}=m_{B_{c1}}$.
Recall that the first of these masses was calculated directly, with the result in Table~\ref{psmesonstatic}.   The last three can be obtained by analysing the appropriate homogeneous Bethe-Salpeter equations using our SPM method.  The results are listed in Table~\ref{SPMparametersMasses}.
% ... Bc* = basis.  Then compute mass-splittings as a function of m ... then extrapolate splittings ... then determine mass.
%
Comparing our predictions with extant lQCD results \cite{Mathur:2018epb}, the mean absolute relative difference is 1.2(0.6)\%.  Looking closer, we find $m_{B_{c0}^\ast}-m_{B_c^\ast}=0.368(13)\,$GeV,
$m_{B_{c1}}-m_{B_c^\ast}=0.474(05)\,$GeV
to be compared with the analogous lQCD results $0.381(20)\,$GeV, $0.405(19)\,$GeV.
The primary differences are that our prediction for $m_{B_c^\ast}$ is 1\% larger than the lQCD result and the axial-vector--vector mass-splitting is 17(6)\% larger.
Considering known empirical masses \cite{Zyla:2020zbs}, the mean $1^{++}-1^{--}$ mass-splitting is $0.416(45)\,$GeV. Our result for $m_{B_{c1}}-m_{B_c^\ast}$ is $15(11)$\% bigger than this.  It may, therefore, be an overestimate.

\begin{table}[t]
\caption{\label{fp0val}
Maximum recoil ($t=0$) value of each transition form factor calculated herein.
Comparisons are provided with other recent analyses:
quark model (QM) \cite{Tran:2018kuv, Issadykov:2018myx};
phenomenology (ph) \cite{Wang:2018duy};
sum rules (SR) \cite{Leljak:2019eyw}
modelling based on perturbative QCD (mpQCD) \cite{Hu:2019qcn};
Salpeter equation (iBS) \cite{Zhou:2019stx};
and lQCD \cite{Colquhoun:2016osw, Harrison:2020gvo}.
As additional context, we list unweighted average values for each of the quantities, evaluated with our prediction excluded (mean-e) and included (mean-i).
}
%
%\begin{tabular}{l|ccc||c}\hline
%%
\begin{tabular*}%{|c|c|c|c|c|c|c|}\hline
{\hsize}
{
l@{\extracolsep{0ptplus1fil}}|
l@{\extracolsep{0ptplus1fil}}
l@{\extracolsep{0ptplus1fil}}
l@{\extracolsep{0ptplus1fil}}
l@{\extracolsep{0ptplus1fil}}
l@{\extracolsep{0ptplus1fil}}}\hline\hline
 & $f_+^{B_c\to\eta_c}\ $ & $A_0^{B_c\to J/\psi}\ $ & $A_1^{B_c\to J/\psi}\ $ & $A_2^{B_c\to J/\psi}\ $ &$V^{B_c\to J/\psi}\ $ \\\hline
herein & $0.63(1)\ $ & $0.57(3)\ $ & $0.55(3)\ $ & $0.56(3)\ $ & $0.84(4)\ $ \\\hline
QM \cite{Tran:2018kuv, Issadykov:2018myx}$\ $ & $0.75\ $  & $0.56\ $ & $0.55\ $ & $0.56\ $ & $0.78\ $ \\
ph \cite{Wang:2018duy}$\ $ & $0.56\ $ & $0.48\ $ & $0.46\ $ & $0.49\ $ & $0.70\ $ \\
SR \cite{Leljak:2019eyw}$\ $ & $0.62(5)\ $ & $0.54(4)\ $ & $0.55(4)\ $ & $0.35(3)\ $ & $0.73(6)\ $ \\
mpQCD \cite{Hu:2019qcn}$\ $ & $0.56(7)\ $ & $0.40(5)\ $ & $0.47(5)\ $ & $0.62(6)\ $ & $0.75(9)\ $ \\
iBS \cite{Zhou:2019stx}$\ $ & $0.41\ $ & $0.46\ $ & $0.48\ $ & $0.54\ $ & $0.63\ $ \\\hline
lQCD \cite{Colquhoun:2016osw}$\ $ & $0.59(1)\ $ & & $0.49(3)\ $ & & $0.70(2)\ $ \\
lQCD \cite{Harrison:2020gvo}$\ $ & & $0.48(3)\ $ & $0.47(3)\ $ & $0.48(8)\ $ & $0.73(7)\ $  \\\hline
mean-e$\ $ & $0.58(11)\ $ & $0.49(6)\ $ & $0.50(4)\ $ & $0.51(9)\ $ & $0.72(5)\ $\\
mean-i$\ $ & $0.59(10)\ $ & $0.50(6)\ $ & $0.50(4)\ $ & $0.51(9)\ $  & $0.73(6)\ $ \\\hline
\end{tabular*}
\end{table}

\begin{figure}[t]
\includegraphics[width=0.42\textwidth]{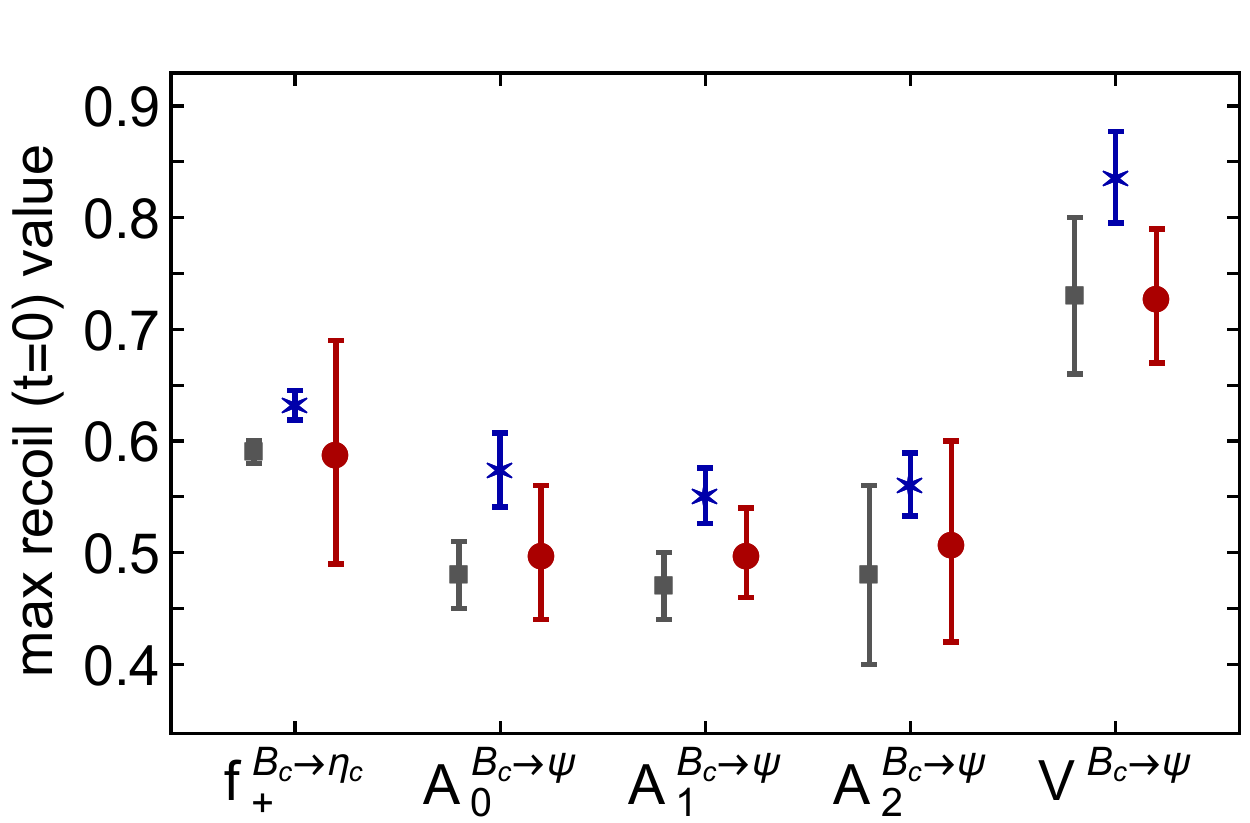}
\caption{\label{Figfp0val}
Predicted values of all transition form factors at the maximum recoil point $(t=0)$ -- blue stars.  Comparisons: lQCD results \cite{Colquhoun:2016osw, Harrison:2020gvo} -- grey squares; and unweighted average of each column in Table~\ref{fp0val} -- red circles.}
\end{figure}

\section{Transition Form Factors: Predictions and Comparisons}
\label{SecPredict}
Our predictions for the $B_c\to \eta_c$ semileptonic transition form factors are given by Eq.\,\eqref{TFfunction} combined with the appropriate masses in Tables~\ref{psmesonstatic}, \ref{SPMparametersMasses} and coefficients listed in Table~\ref{SPMparameters}.  The maximum recoil $(t=0)$ value of each form factor is listed in Table~\ref{fp0val}, compared with recent continuum and lattice estimates.  Aspects of the information in Table~\ref{fp0val} are depicted in Fig.\,\ref{Figfp0val}.  Evidently, different approaches produce a range of values for $f_+(0)$; nevertheless, all values fall within $\lesssim 20$\% of the mean.

\begin{figure}[t]
\vspace*{2ex}

\leftline{\hspace*{0.5em}{\large{\textsf{A}}}}
\vspace*{-3ex}
\includegraphics[width=0.42\textwidth]{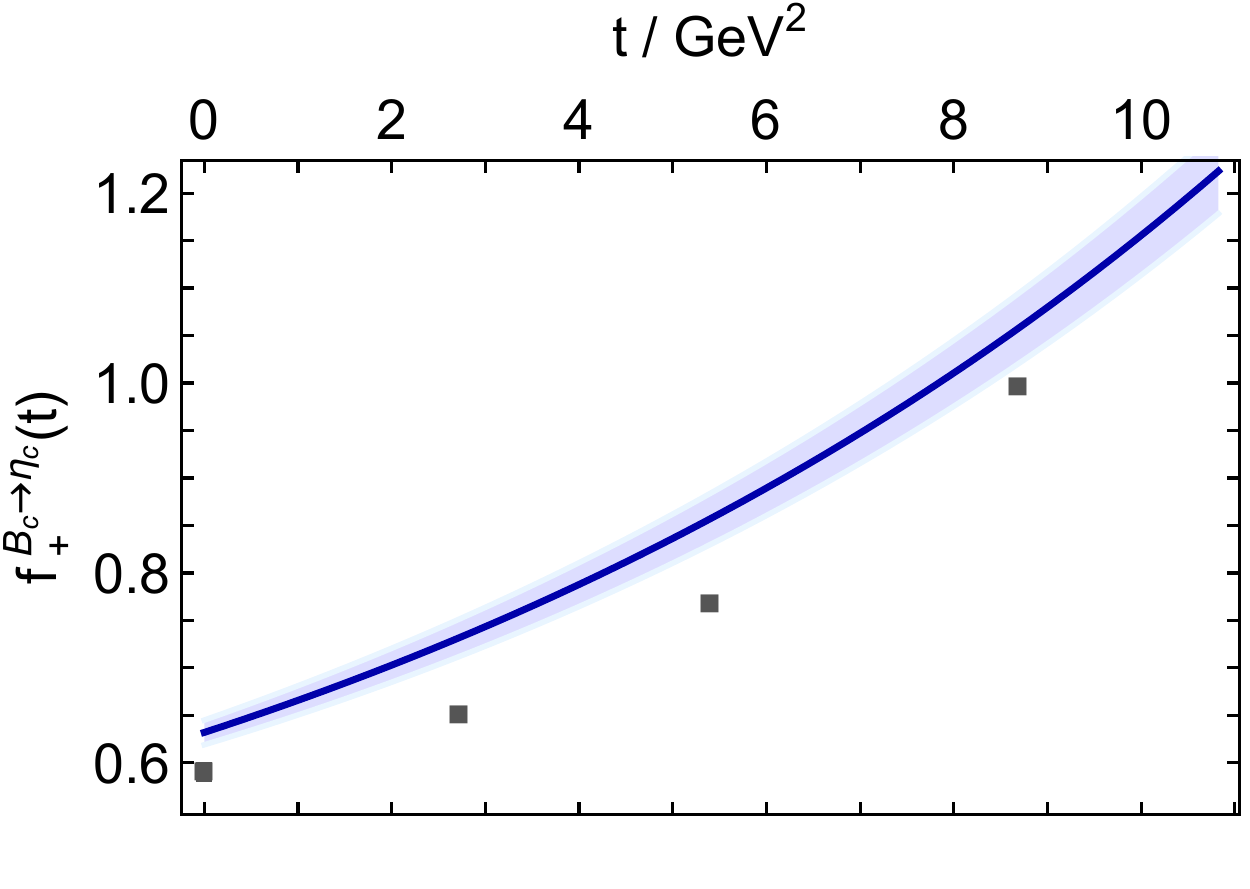}
\vspace*{-1ex}

\leftline{\hspace*{0.5em}{\large{\textsf{B}}}}
\vspace*{-3ex}
\includegraphics[width=0.42\textwidth]{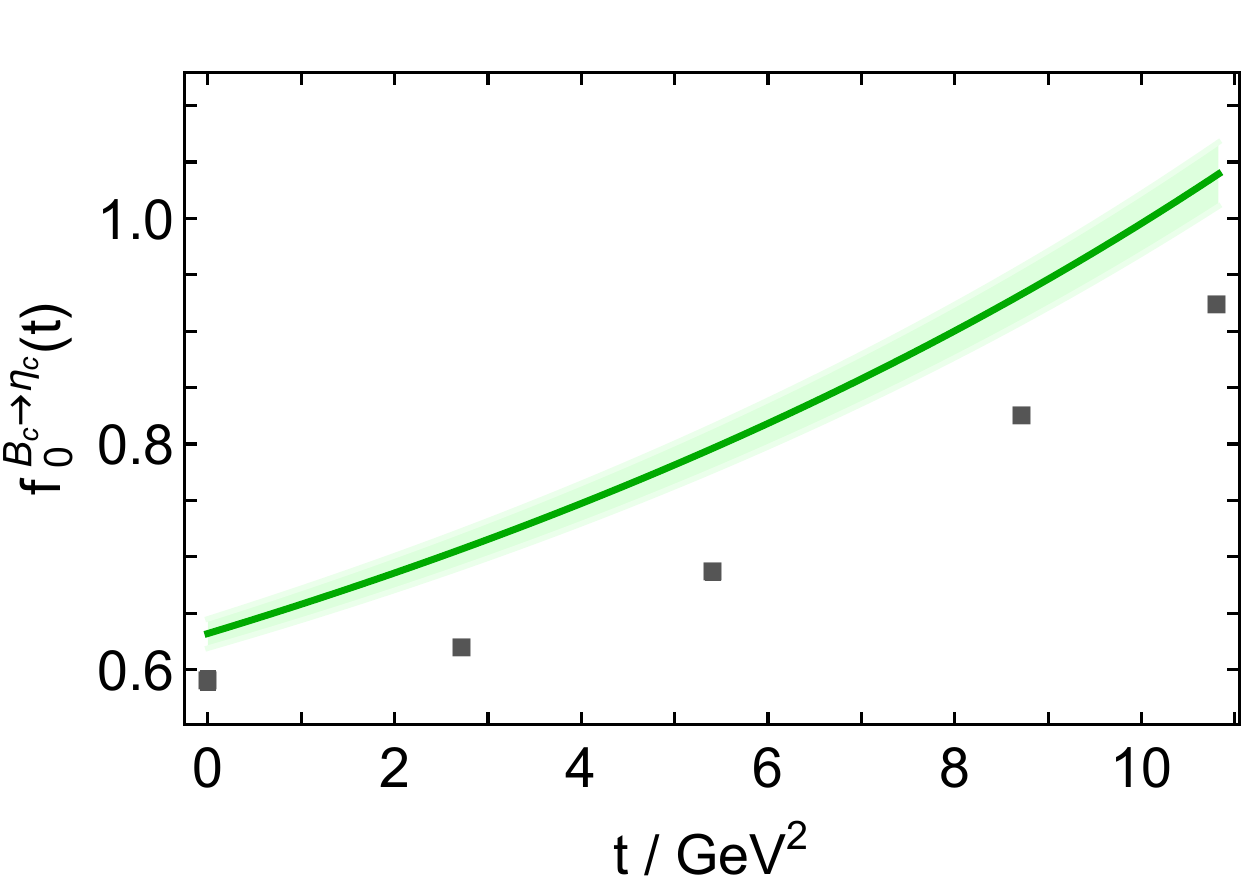}

\caption{\label{FigBcetac}
Predicted $B_c\to \eta_c$ semileptonic transition form factors.
%\emph{Upper panel}\,--\,{\sf A}: $f_+(t)$; and \emph{lower panel}\,--\,{\sf B}: $f_0(t)$.
%
(The shaded bands surrounding each curve express the SPM uncertainty, determined as discussed in the paragraph before that containing Eq.\,\eqref{mBcSPM}.)
The points in both panels are preliminary lQCD results from Ref.\,\cite{Colquhoun:2016osw}.
}
\end{figure}

\begin{table}[t]
\caption{\label{TabBF}
Branching fractions calculated using our predictions for the semileptonic transition form factors in Eqs.\,\eqref{dGdt}\,--\,\eqref{Hpsi} and empirical lepton and meson masses: ({\sf A}) -- $B_c\to \eta_c$; and (B) -- $B_c\to J/\psi$.
Two uncertainties are listed with our results: first -- $1\sigma$ SPM uncertainty; second -- from error on $|V_{cb}|$.
Column~3 reports the ratio of the first two columns: $|V_{cb}|$ cancels.
Comparisons are provided with other analyses:
quark model (QM) \cite{Tran:2018kuv, Issadykov:2018myx};
phenomenology (ph) \cite{Wang:2018duy};
sum rules (SR) \cite{Leljak:2019eyw}
modelling based on perturbative QCD (mpQCD) \cite{Hu:2019qcn};
Salpeter equation (iBS) \cite{Zhou:2019stx};
and lQCD \cite{Harrison:2020nrv, Harrison:2020gvo}.  (No lQCD results are available for inclusion in Panel {\sf A}.)
As additional context, we list an unweighted average value for each quantity, evaluated with our prediction excluded (mean-e) and included (mean-i).
Branching fractions are to be multiplied by $10^{-3}$.
}
%
%\begin{tabular}{l|ccc||c}\hline
%%
\begin{tabular*}%{|c|c|c|c|c|c|c|}\hline
{\hsize}
{
l@{\extracolsep{0ptplus1fil}}|
l@{\extracolsep{0ptplus1fil}}
l@{\extracolsep{0ptplus1fil}}|
l@{\extracolsep{0ptplus1fil}}}\hline\hline
{\sf A}
& ${\mathpzc B}_{B_c\to\eta_c\mu \nu_\mu}\ $
& ${\mathpzc B}_{B_c\to\eta_c\tau \nu_\tau}\ $
& $R_{\eta_c}\ $ \\\hline
herein & $8.10 \, (45)\,(55)\ $ & $2.54(10)(17)\ $ & $0.31(2)\ $ \\\hline
QM \cite{Tran:2018kuv, Issadykov:2018myx}$\ $ & $9.5(1.9)\ $ & $2.4(0.5)\ $ & $0.25(7)\ $ \\
ph \cite{Wang:2018duy}$\ $ & $6.6(0.2)\ $ & $\ $ & $0.31(1)\ $ \\
SR \cite{Leljak:2019eyw}$\ $ & $8.2(1.2)\ $ & $2.6(0.6)\ $ & $0.32(2)\ $ \\
mpQCD \cite{Hu:2019qcn}$\ $ & $7.8(1.7)\ $ & $2.4(0.4)\ $ & $0.31(1)\ $ \\
iBS \cite{Zhou:2019stx}$\ $ & $5.3(2.2)\ $ & $2.2(0.7)\ $ & $0.38(4)\ $ \\
\hline
mean-e$\ $ &$7.5(1.6)\ $ & $2.4(0.2)\ $ & $0.31(4)\ $ \\
mean-i$\ $  & $7.6(1.5)\ $ & $2.4(0.2)\ $ & $0.31(4)\ $  \\\hline
\end{tabular*}
\medskip

\begin{tabular*}%{|c|c|c|c|c|c|c|}\hline
{\hsize}
{
l@{\extracolsep{0ptplus1fil}}|
l@{\extracolsep{0ptplus1fil}}
l@{\extracolsep{0ptplus1fil}}|
l@{\extracolsep{0ptplus1fil}}}\hline\hline
{\sf B}
& ${\mathpzc B}_{B_c\to J/\psi \mu \nu_\mu}\ $
& ${\mathpzc B}_{B_c\to J/\psi \tau \nu_\tau}\ $
& $R_{J/\psi}\ $ \\\hline
herein & $17.2 \, (1.9)\,(1.2)\ $ & $4.17(66)(28)\ $ & $0.24(5)\ $ \\\hline
QM \cite{Tran:2018kuv, Issadykov:2018myx}$\ $ & $16.7(3.3)\ $ & $4.0(0.8)\ $ & $0.24(7)\ $ \\
ph \cite{Wang:2018duy}$\ $ & $14.4(0.2)\ $ & $\ $ & $0.26(1)\ $ \\
SR \cite{Leljak:2019eyw}$\ $ & $22.4(5.3)\ $ & $5.3(1.5)\ $ & $0.23(1)\ $ \\
mpQCD \cite{Hu:2019qcn}$\ $ & $14.1(2.5)\ $ & $3.8(0.6)\ $ & $0.27(1)\ $ \\
iBS \cite{Zhou:2019stx}$\ $ & $16.2(0.5)\ $ & $4.3(0.1)\ $ & $0.27(1)\ $ \\
\hline
lQCD \cite{Harrison:2020nrv, Harrison:2020gvo}$\ $ & $15.0(1.1)(1.0)\ $ &  & $0.258(4)\ $ \\
\hline
mean-e$\ $ &$16.5(3.1)\ $ & $4.4(0.7)\ $ & $0.25(2)\ $ \\
mean-i$\ $  & $16.6(2.8)\ $ & $4.3(0.6)\ $ & $0.25(2)\ $  \\\hline
\end{tabular*}
\end{table}

Our $B_c\to \eta_c$ transition form factors are drawn in Figs.\,\ref{FigBcetac}.  The difference between these predictions and the preliminary lQCD results reported in Ref.\,\cite{Colquhoun:2016osw} is 10(3)\%, with the lQCD values lying uniformly below our curves.  No further information on $B_c\to \eta_c$ is currently available from lQCD.  Here, therefore, the interpolations we provide for our calculated transition form factors can be valuable in analysing future experimental data on the related transitions.

Working with our predictions and using Eqs.\,\eqref{dGdt}\,--\,\eqref{Hpsi} evaluated with empirical lepton and meson masses, we obtain the $B_c\to \eta_c$ branching fractions reported in Table~\ref{TabBF}A.  Our results match well with other contemporary estimates.  %No comparable lQCD results are available in this case.

Similarly, our predictions for the $B_c\to J/\psi$ semileptonic transition form factors are given by Eq.\,\eqref{TFfunction} combined with the appropriate masses in Tables~\ref{psmesonstatic}, \ref{SPMparametersMasses} and coefficients listed in Table~\ref{SPMparameters}.  The maximum recoil $(t=0)$ value of each form factor is listed in Table~\ref{fp0val}, compared with recent continuum and lattice estimates.  Once again, as highlighted by Fig.\,\ref{Figfp0val}, different approaches produce a range of $t=0$ form factor values; but there is no significant tension, with all values falling within $\lesssim 15$\% of their respective means.

\begin{figure*}[!t]
\begin{center}
\begin{tabular}{lr}
{\large\sf A} & {\large\sf B} \\[-4ex]
\includegraphics[clip,width=0.42\linewidth]{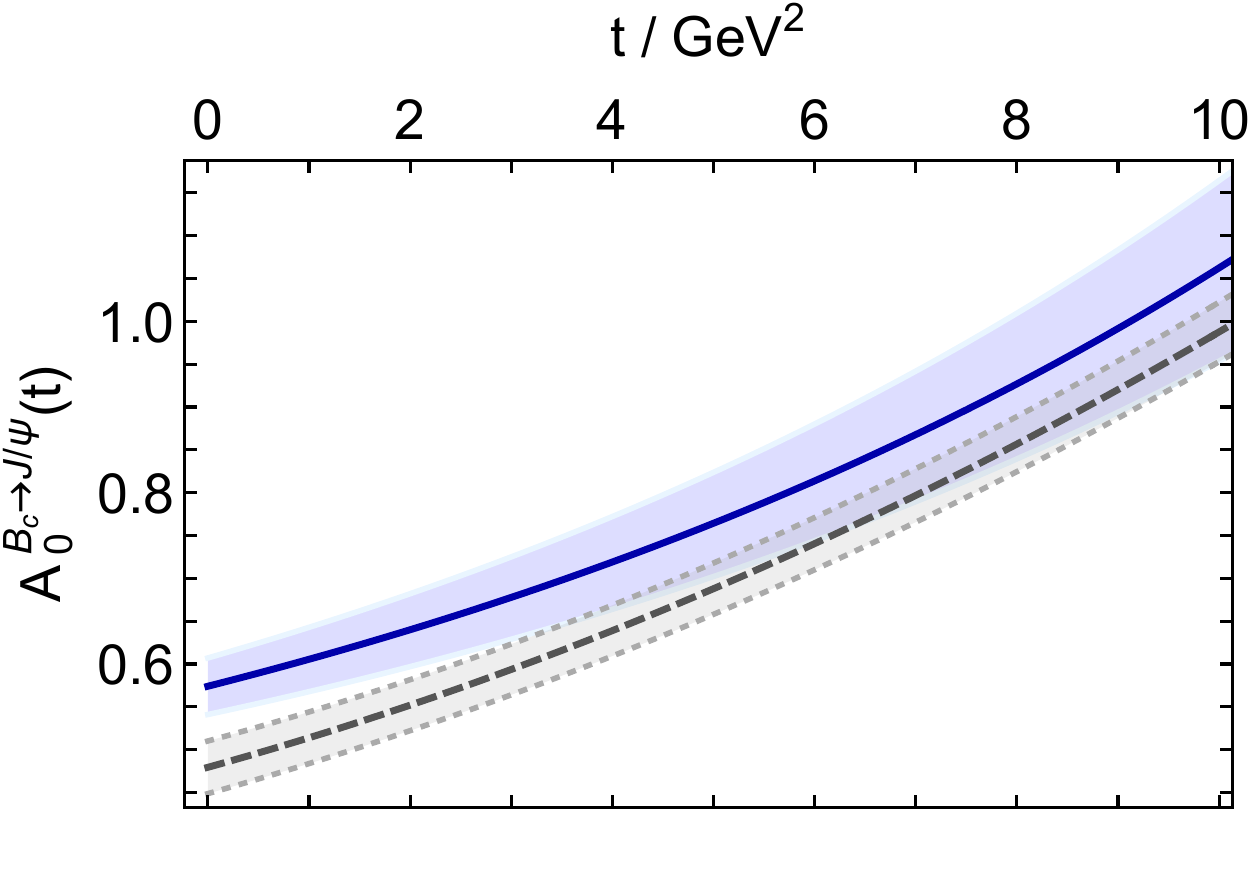}\hspace*{2ex } &
\includegraphics[clip,width=0.42\linewidth]{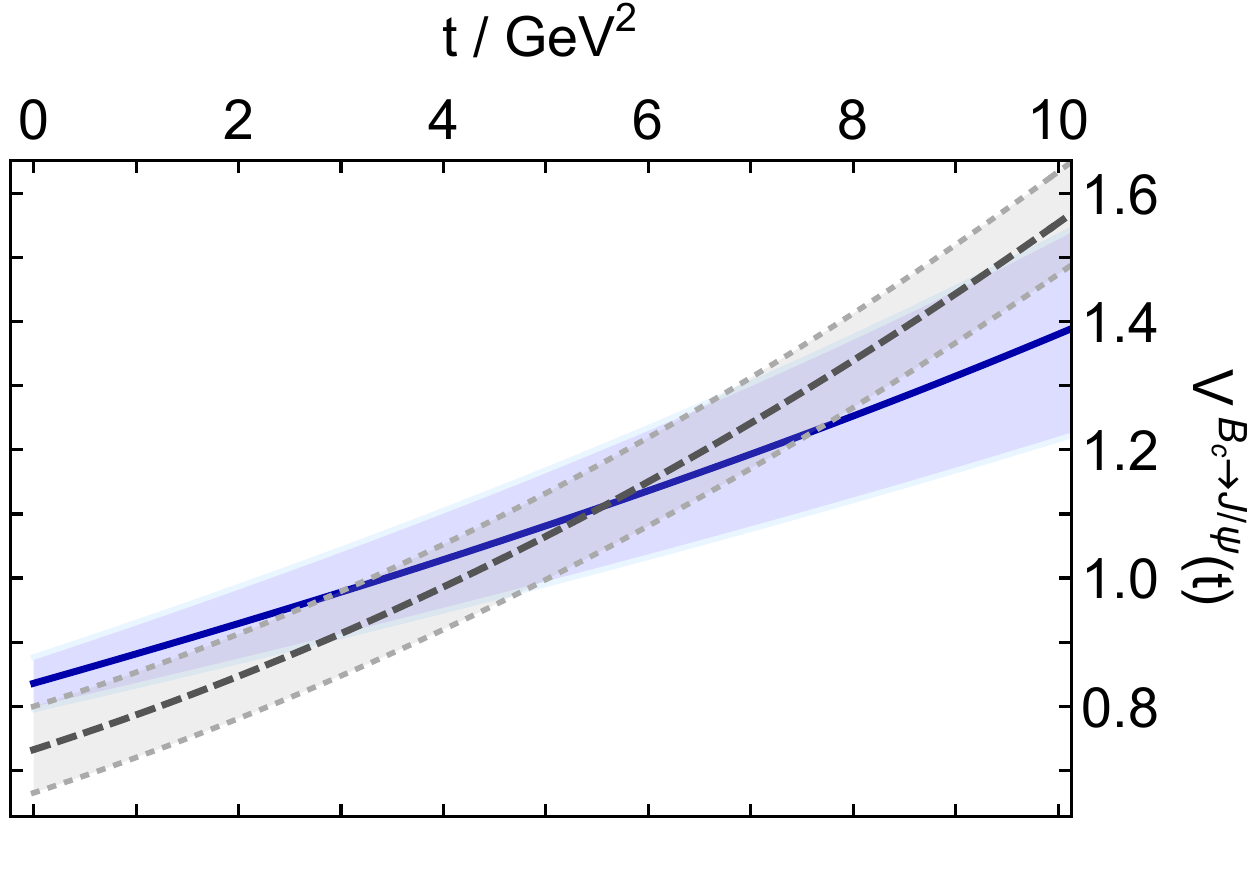}\vspace*{-0ex}
\end{tabular}\vspace*{-3ex}
\begin{tabular}{lr}
{\large\sf C} & {\large\sf D} \\[-4ex]
\includegraphics[clip,width=0.42\linewidth]{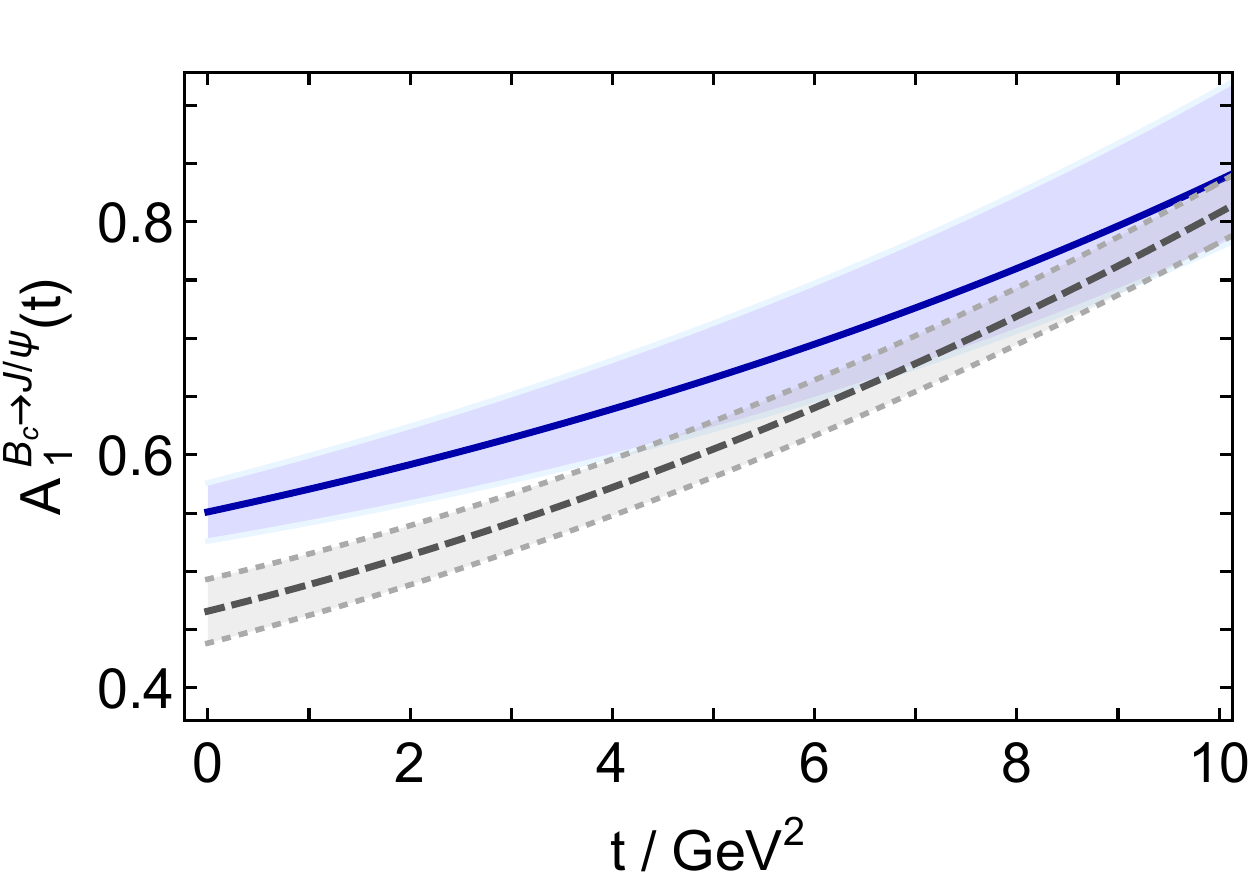}\hspace*{2ex } &
\includegraphics[clip,width=0.42\linewidth]{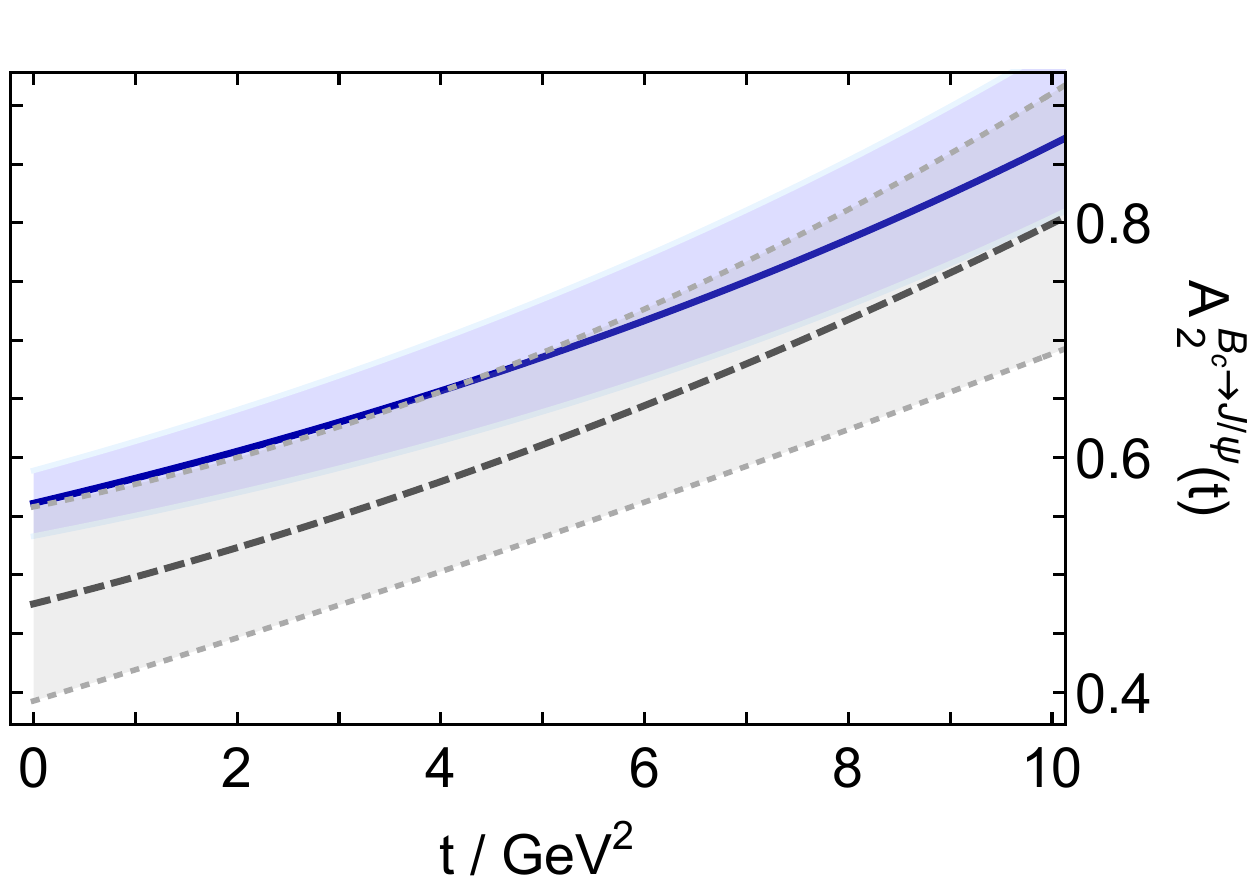}\vspace*{-1ex}
\end{tabular}
\end{center}
\caption{\label{elastic}
Predicted $B_c\to J/\psi$ semileptonic transition form factors -- solid blue curves within like-coloured bands, which express the $1\sigma$ uncertainty on our SPM results.
Comparable lQCD results from Ref.\,\cite{Harrison:2020gvo} -- dashed grey curves within like-coloured bands.
}
\end{figure*}

Our $B_c\to J/\psi$ transition form factors are depicted in Figs.\,\ref{elastic}.  Comparing with lQCD results \cite{Harrison:2020gvo}, despite minor qualitative differences, most notably concerning $V^{B_c\to J/\psi}(t)$ in Fig.\,\ref{elastic}B, there is semi-quantitative agreement.
The interpolations we provide for our calculated transition form factors could be used to reduce a dominant systematic error in the extraction of $R_{J/\psi}$ from experiment \cite{Aaij:2017tyk}, paving the way to improved precision and a more stringent test of the Standard Model.

Using Eqs.\,\eqref{dGdt}\,--\,\eqref{Hpsi} evaluated with empirical lepton and meson masses and our predictions in Figs.\,\ref{elastic}, we obtain the $B_c\to J/\psi$ branching fractions reported in Table~\ref{TabBF}B.  Our results accord well with other contemporary estimates.

\section{Conclusions and Perspectives}
We employed a systematic, symmetry-preserving approach to the continuum strong-interaction bound-state problem in the Standard Model to calculate the semileptonic $B_c \to \eta_c, J/\psi$ transition form factors on the entire physical kinematic domain.  The framework [Sec.\,\ref{SecRL}] has been used successfully to unify the properties of mesons and baryons with $0-3$ heavy-quarks; and from this foundation, we arrived at an array of parameter-free predictions, including the branching fractions  ${\mathpzc B}_{B_c\to\eta_c {\mathpzc l} \nu_{\mathpzc l}}$, ${\mathpzc B}_{B_c\to J/\psi  {\mathpzc l} \nu_{\mathpzc l}}\ $, ${\mathpzc l}=\mu,\tau$ [Sec.\,\ref{SecPredict}].

A key result of our analysis is highlighted by Fig.\,\ref{RJpsi}.  Namely, contemporary Standard Model calculations of the ratio $R_{J/\psi}$ in Eq.\,\eqref{eqRjpsi} are in agreement.  Combined via the mean of their central values, they produce $R_{J/\psi}=0.253(16)$, which is approximately $2\sigma$ below the empirical result reported in Ref.\,\cite{Aaij:2017tyk}.  If subsequent, precision experiments do not lead to a substantially lower central value, then one may conclude that lepton flavour universality is violated in semileptonic $B_c\to J/\psi$ decays.  However, the precision of existing empirical information is insufficient to support such a claim.  Moreover, a compelling case could only be compiled by including information on semileptonic $B_c\to \eta_c$ decays.  We predict $R_{\eta_c}=0.313(22)$; and the mean obtained from modern continuum analyses is $0.31(4)$ [Table~\ref{TabBF}A].

Natural extensions of this work include kindred analyses of $b\to c$ transitions in the semileptonic decays of $B_{(s)}$ mesons with $D_{(s)}^{(\ast)}$ mesons in the final-state.
Existing surveys of Standard Model theory estimates of the  ``$R$'' ratios associated with these additional processes yield values similar to those discussed herein, with the result for the pseudoscalar-meson final-state being $\sim 15$\% greater than that for the vector-meson final-state \cite{Hu:2019bdf, Gambino:2020jvv}.
%% average from 1st ref Ds/Ds*  0.32(2)  &  0.29(5)
%% Table 3  D/D* 0.299(3) & 0.258(5)
%
Augmenting such analyses via the parameter-free unification of the results obtained herein with predictions for these other ``$R$'' ratios should serve to increase confidence in Standard Model predictions and strengthen any case for or against lepton flavour universality in Nature.
Furthermore, one could expand the coverage of our study to include a wider range of measurable quantities \cite{Tran:2018kuv, Zhang:2020dla}, providing additional benchmarks for Standard Model tests in $B_c$ decays.

\smallskip

\noindent\textbf{Acknowledgments}\,---\,%We are grateful for insightful comments from .
%\acknowledgments
%
We are grateful for constructive comments from X.-W.~Kang and Z.-N.~Xu; and to H.-S.~Zong for friendship and guidance until his untimely death before this study was completed.
Work supported by:
Nanjing University Innovation Programme for PhD candidates;
National Natural Science Foundation of China (under Grant No.\,11805097);
Jiangsu Provincial Natural Science Foundation of China (under Grant No.\,BK20180323);
and
STRONG-2020 ``The strong interaction at the frontier of knowledge: fundamental research and applications'', which received funding from the European Union’s Horizon 2020 research and innovation programme under grant agreement No 824093.
%
%Chinese Ministry of Education, under the \emph{International Distinguished Professor} programme.


%\pagebreak

%%\medskip

%%\noindent\textbf{References}\,---\,%We are grateful for insightful comments from .
%%\bibliographystyle{elsarticle-num-names}
%%\bibliography{../../../CollectedBiB}

\begin{thebibliography}{70}
\providecommand{\natexlab}[1]{#1}
\providecommand{\url}[1]{\texttt{#1}}
\providecommand{\urlprefix}{URL }
\expandafter\ifx\csname urlstyle\endcsname\relax
  \providecommand{\doi}[1]{doi:\discretionary{}{}{}#1}\else
  \providecommand{\doi}[1]{doi:\discretionary{}{}{}\begingroup
  \urlstyle{rm}\url{#1}\endgroup}\fi
\providecommand{\bibinfo}[2]{#2}

\bibitem[{Abe et~al.(1998)}]{Abe:1998fb}
\bibinfo{author}{F.~Abe}, et~al., \bibinfo{title}{{Observation of $B_c$ mesons
  in $p\bar{p}$ collisions at $\sqrt{s} = 1.8$ TeV}}, \bibinfo{journal}{Phys.
  Rev. D} \bibinfo{volume}{58} (\bibinfo{year}{1998}) \bibinfo{pages}{112004}.

\bibitem[{Zyla et~al.(2020)}]{Zyla:2020zbs}
\bibinfo{author}{P.~Zyla}, et~al., \bibinfo{title}{{Review of Particle
  Physics}}, \bibinfo{journal}{PTEP} \bibinfo{volume}{2020}
  (\bibinfo{year}{2020}) \bibinfo{pages}{083C01}.

\bibitem[{Barik et~al.(2009)Barik, Naimuddin, Dash, and Kar}]{Barik:2009zz}
\bibinfo{author}{N.~Barik}, \bibinfo{author}{S.~Naimuddin},
  \bibinfo{author}{P.~C. Dash}, \bibinfo{author}{S.~Kar},
  \bibinfo{title}{{Semileptonic decays of the $B_c$ meson}},
  \bibinfo{journal}{Phys. Rev. D} \bibinfo{volume}{80} (\bibinfo{year}{2009})
  \bibinfo{pages}{074005}.

\bibitem[{Zhang et~al.(2021)Zhang, Kang, Guo, Dai, Luo, and
  Wang}]{Zhang:2020dla}
\bibinfo{author}{L.~Zhang}, \bibinfo{author}{X.-W. Kang},
  \bibinfo{author}{X.-H. Guo}, \bibinfo{author}{L.-Y. Dai},
  \bibinfo{author}{T.~Luo}, \bibinfo{author}{C.~Wang}, \bibinfo{title}{{A
  comprehensive study on the semileptonic decay of heavy flavor mesons}},
  \bibinfo{journal}{JHEP} \bibinfo{volume}{02} (\bibinfo{year}{2021})
  \bibinfo{pages}{179}.

\bibitem[{Xu et~al.(2021)Xu, Cui, Roberts, and Xu}]{Xu:2021iwv}
\bibinfo{author}{Z.-N. Xu}, \bibinfo{author}{Z.-F. Cui}, \bibinfo{author}{C.~D.
  Roberts}, \bibinfo{author}{C.~Xu}, \bibinfo{title}{{Heavy+light pseudoscalar
  meson semileptonic transitions -- arXiv:2103.15964 [hep-ph]}} .

\bibitem[{Scora and Isgur(1995)}]{Scora:1995ty}
\bibinfo{author}{D.~Scora}, \bibinfo{author}{N.~Isgur},
  \bibinfo{title}{{Semileptonic meson decays in the quark model: An update}},
  \bibinfo{journal}{Phys. Rev. D} \bibinfo{volume}{52} (\bibinfo{year}{1995})
  \bibinfo{pages}{2783--2812}.

\bibitem[{Gouz et~al.(2004)Gouz, Kiselev, Likhoded, Romanovsky, and
  Yushchenko}]{Gouz:2002kk}
\bibinfo{author}{I.~P. Gouz}, \bibinfo{author}{V.~V. Kiselev},
  \bibinfo{author}{A.~K. Likhoded}, \bibinfo{author}{V.~I. Romanovsky},
  \bibinfo{author}{O.~P. Yushchenko}, \bibinfo{title}{{Prospects for the $B_c$
  studies at LHCb}}, \bibinfo{journal}{Phys. Atom. Nucl.} \bibinfo{volume}{67}
  (\bibinfo{year}{2004}) \bibinfo{pages}{1559--1570}.

\bibitem[{Aaij et~al.(2018{\natexlab{a}})}]{Aaij:2017tyk}
\bibinfo{author}{R.~Aaij}, et~al., \bibinfo{title}{{Measurement of the ratio of
  branching fractions
  $\mathcal{B}(B_c^+\,\to\,J/\psi\tau^+\nu_\tau)$/$\mathcal{B}(B_c^+\,\to\,J/\psi\mu^+\nu_\mu)$}},
  \bibinfo{journal}{Phys. Rev. Lett.}
  \bibinfo{volume}{120}~(\bibinfo{number}{12})
  (\bibinfo{year}{2018}{\natexlab{a}}) \bibinfo{pages}{121801}.

\bibitem[{Harrison et~al.(2020{\natexlab{a}})Harrison, Davies, and
  Lytle}]{Harrison:2020nrv}
\bibinfo{author}{J.~Harrison}, \bibinfo{author}{C.~T.~H. Davies},
  \bibinfo{author}{A.~Lytle}, \bibinfo{title}{{$R(J/\psi)$ and $B_c^-
  \rightarrow J/\psi \ell^-\bar{\nu}_\ell$ Lepton Flavor Universality Violating
  Observables from Lattice QCD}}, \bibinfo{journal}{Phys. Rev. Lett.}
  \bibinfo{volume}{125}~(\bibinfo{number}{22})
  (\bibinfo{year}{2020}{\natexlab{a}}) \bibinfo{pages}{222003}.

\bibitem[{Harrison et~al.(2020{\natexlab{b}})Harrison, Davies, and
  Lytle}]{Harrison:2020gvo}
\bibinfo{author}{J.~Harrison}, \bibinfo{author}{C.~T.~H. Davies},
  \bibinfo{author}{A.~Lytle}, \bibinfo{title}{{$B_c \rightarrow J/\psi$ form
  factors for the full $q^2$ range from lattice QCD}}, \bibinfo{journal}{Phys.
  Rev. D} \bibinfo{volume}{102} (\bibinfo{year}{2020}{\natexlab{b}})
  \bibinfo{pages}{094518}.

\bibitem[{Tran et~al.(2018)Tran, Ivanov, K\"orner, and
  Santorelli}]{Tran:2018kuv}
\bibinfo{author}{C.-T. Tran}, \bibinfo{author}{M.~A. Ivanov},
  \bibinfo{author}{J.~G. K\"orner}, \bibinfo{author}{P.~Santorelli},
  \bibinfo{title}{{Implications of new physics in the decays $B_c \to
  (J/\psi,\eta_c)\tau\nu$}}, \bibinfo{journal}{Phys. Rev. D}
  \bibinfo{volume}{97} (\bibinfo{year}{2018}) \bibinfo{pages}{054014}.

\bibitem[{Issadykov and Ivanov(2018)}]{Issadykov:2018myx}
\bibinfo{author}{A.~Issadykov}, \bibinfo{author}{M.~A. Ivanov},
  \bibinfo{title}{{The decays $B_{c}\to J/\psi+\bar\ell\nu_\ell$ and $B_{c}\to
  J/\psi + \pi(K)$ in covariant confined quark model}}, \bibinfo{journal}{Phys.
  Lett. B} \bibinfo{volume}{783} (\bibinfo{year}{2018})
  \bibinfo{pages}{178--182}.

\bibitem[{Wang and Zhu(2019)}]{Wang:2018duy}
\bibinfo{author}{W.~Wang}, \bibinfo{author}{R.~Zhu}, \bibinfo{title}{{Model
  independent investigation of the $R_{J/\psi,\eta_c}$ and ratios of decay
  widths of semileptonic $B_c$ decays into a P-wave charmonium}},
  \bibinfo{journal}{Int. J. Mod. Phys. A}
  \bibinfo{volume}{34}~(\bibinfo{number}{31}) (\bibinfo{year}{2019})
  \bibinfo{pages}{1950195}.

\bibitem[{Leljak et~al.(2019)Leljak, Melic, and Patra}]{Leljak:2019eyw}
\bibinfo{author}{D.~Leljak}, \bibinfo{author}{B.~Melic},
  \bibinfo{author}{M.~Patra}, \bibinfo{title}{{On lepton flavour universality
  in semileptonic $B_c \to \eta_c, J/\psi$ decays}}, \bibinfo{journal}{JHEP}
  \bibinfo{volume}{05} (\bibinfo{year}{2019}) \bibinfo{pages}{094}.

\bibitem[{Hu et~al.(2020{\natexlab{a}})Hu, Jin, and Xiao}]{Hu:2019qcn}
\bibinfo{author}{X.-Q. Hu}, \bibinfo{author}{S.-P. Jin}, \bibinfo{author}{Z.-J.
  Xiao}, \bibinfo{title}{{Semileptonic decays $B_c \to (\eta_c,J/\psi) l
  \bar{\nu}_l $ in the ``PQCD + Lattice'' approach}}, \bibinfo{journal}{Chin.
  Phys. C} \bibinfo{volume}{44}~(\bibinfo{number}{2})
  (\bibinfo{year}{2020}{\natexlab{a}}) \bibinfo{pages}{023104}.

\bibitem[{Zhou et~al.(2020)Zhou, Wang, Jiang, Tan, Li, and Wang}]{Zhou:2019stx}
\bibinfo{author}{T.~Zhou}, \bibinfo{author}{T.~Wang},
  \bibinfo{author}{Y.~Jiang}, \bibinfo{author}{X.-Z. Tan},
  \bibinfo{author}{G.~Li}, \bibinfo{author}{G.-L. Wang},
  \bibinfo{title}{{Relativistic calculations of $R(D^{(*)})$, $R(D^{(*)}_s)$,
  $R(\eta_c)$ and $R(J/\psi)$}}, \bibinfo{journal}{Int. J. Mod. Phys. A}
  \bibinfo{volume}{35}~(\bibinfo{number}{17}) (\bibinfo{year}{2020})
  \bibinfo{pages}{2050076}.

\bibitem[{Lees et~al.(2013)}]{Lees:2013uzd}
\bibinfo{author}{J.~P. Lees}, et~al., \bibinfo{title}{{Measurement of an Excess
  of $\bar{B} \to D^{(*)}\tau^- \bar{\nu}_\tau$ Decays and Implications for
  Charged Higgs Bosons}}, \bibinfo{journal}{Phys. Rev. D}
  \bibinfo{volume}{88}~(\bibinfo{number}{7}) (\bibinfo{year}{2013})
  \bibinfo{pages}{072012}.

\bibitem[{Huschle et~al.(2015)}]{Huschle:2015rga}
\bibinfo{author}{M.~Huschle}, et~al., \bibinfo{title}{{Measurement of the
  branching ratio of $\bar{B} \to D^{(\ast)} \tau^- \bar{\nu}_\tau$ relative to
  $\bar{B} \to D^{(\ast)} \ell^- \bar{\nu}_\ell$ decays with hadronic tagging
  at Belle}}, \bibinfo{journal}{Phys. Rev. D}
  \bibinfo{volume}{92}~(\bibinfo{number}{7}) (\bibinfo{year}{2015})
  \bibinfo{pages}{072014}.

\bibitem[{Aaij et~al.(2015)}]{Aaij:2015yra}
\bibinfo{author}{R.~Aaij}, et~al., \bibinfo{title}{{Measurement of the ratio of
  branching fractions $\mathcal{B}(\bar{B}^0 \to
  D^{*+}\tau^{-}\bar{\nu}_{\tau})/ \mathcal{B}(\bar{B}^0 \to
  D^{*+}\mu^{-}\bar{\nu}_{\mu})$}}, \bibinfo{journal}{Phys. Rev. Lett.}
  \bibinfo{volume}{115}~(\bibinfo{number}{11}) (\bibinfo{year}{2015})
  \bibinfo{pages}{111803}, \bibinfo{note}{[Erratum: Phys. Rev. Lett.
  \textbf{115}, 159901 (2015)]}.

\bibitem[{Sato et~al.(2016)}]{Sato:2016svk}
\bibinfo{author}{Y.~Sato}, et~al., \bibinfo{title}{{Measurement of the
  branching ratio of $\bar{B}^0 \rightarrow D^{*+} \tau^- \bar{\nu}_{\tau}$
  relative to $\bar{B}^0 \rightarrow D^{*+} \ell^- \bar{\nu}_{\ell}$ decays
  with a semileptonic tagging method}}, \bibinfo{journal}{Phys. Rev. D}
  \bibinfo{volume}{94}~(\bibinfo{number}{7}) (\bibinfo{year}{2016})
  \bibinfo{pages}{072007}.

\bibitem[{Hirose et~al.(2017)}]{Hirose:2016wfn}
\bibinfo{author}{S.~Hirose}, et~al., \bibinfo{title}{{Measurement of the $\tau$
  lepton polarization and $R(D^*)$ in the decay $\bar{B} \to D^* \tau^-
  \bar{\nu}_\tau$}}, \bibinfo{journal}{Phys. Rev. Lett.}
  \bibinfo{volume}{118}~(\bibinfo{number}{21}) (\bibinfo{year}{2017})
  \bibinfo{pages}{211801}.

\bibitem[{Aaij et~al.(2018{\natexlab{b}})}]{Aaij:2017uff}
\bibinfo{author}{R.~Aaij}, et~al., \bibinfo{title}{{Measurement of the ratio of
  the $B^0 \to D^{*-} \tau^+ \nu_{\tau}$ and $B^0 \to D^{*-} \mu^+ \nu_{\mu}$
  branching fractions using three-prong $\tau$-lepton decays}},
  \bibinfo{journal}{Phys. Rev. Lett.}
  \bibinfo{volume}{120}~(\bibinfo{number}{17})
  (\bibinfo{year}{2018}{\natexlab{b}}) \bibinfo{pages}{171802}.

\bibitem[{Aaij et~al.(2021)}]{Aaij:2021vac}
\bibinfo{author}{R.~Aaij}, et~al., \bibinfo{title}{{Test of lepton universality
  in beauty-quark decays}} .

\bibitem[{Berns and Lamm(2018)}]{Berns:2018vpl}
\bibinfo{author}{A.~Berns}, \bibinfo{author}{H.~Lamm},
  \bibinfo{title}{{Model-Independent Prediction of $R(\eta_c)$}},
  \bibinfo{journal}{JHEP} \bibinfo{volume}{12} (\bibinfo{year}{2018})
  \bibinfo{pages}{114}.

\bibitem[{Cohen et~al.(2019)Cohen, Lamm, and Lebed}]{Cohen:2019zev}
\bibinfo{author}{T.~D. Cohen}, \bibinfo{author}{H.~Lamm},
  \bibinfo{author}{R.~F. Lebed}, \bibinfo{title}{{Precision Model-Independent
  Bounds from Global Analysis of $b \to c \ell \nu$ Form Factors}},
  \bibinfo{journal}{Phys. Rev. D} \bibinfo{volume}{100} (\bibinfo{year}{2019})
  \bibinfo{pages}{094503}.

\bibitem[{Colquhoun et~al.(2016)Colquhoun, Davies, Koponen, Lytle, and
  McNeile}]{Colquhoun:2016osw}
\bibinfo{author}{B.~Colquhoun}, \bibinfo{author}{C.~Davies},
  \bibinfo{author}{J.~Koponen}, \bibinfo{author}{A.~Lytle},
  \bibinfo{author}{C.~McNeile}, \bibinfo{title}{{$B_c$ decays from highly
  improved staggered quarks and NRQCD}}, \bibinfo{journal}{PoS}
  \bibinfo{volume}{LATTICE2016} (\bibinfo{year}{2016}) \bibinfo{pages}{281}.

\bibitem[{Horn and Roberts(2016)}]{Horn:2016rip}
\bibinfo{author}{T.~Horn}, \bibinfo{author}{C.~D. Roberts},
  \bibinfo{title}{{The pion: an enigma within the Standard Model}},
  \bibinfo{journal}{J. Phys. G.} \bibinfo{volume}{43} (\bibinfo{year}{2016})
  \bibinfo{pages}{073001}.

\bibitem[{Eichmann et~al.(2016)Eichmann, Sanchis-Alepuz, Williams, Alkofer, and
  Fischer}]{Eichmann:2016yit}
\bibinfo{author}{G.~Eichmann}, \bibinfo{author}{H.~Sanchis-Alepuz},
  \bibinfo{author}{R.~Williams}, \bibinfo{author}{R.~Alkofer},
  \bibinfo{author}{C.~S. Fischer}, \bibinfo{title}{{Baryons as relativistic
  three-quark bound states}}, \bibinfo{journal}{Prog. Part. Nucl. Phys.}
  \bibinfo{volume}{91} (\bibinfo{year}{2016}) \bibinfo{pages}{1--100}.

\bibitem[{Fischer(2019)}]{Fischer:2018sdj}
\bibinfo{author}{C.~S. Fischer}, \bibinfo{title}{{QCD at finite temperature and
  chemical potential from Dyson--Schwinger equations}}, \bibinfo{journal}{Prog.
  Part. Nucl. Phys.} \bibinfo{volume}{105} (\bibinfo{year}{2019})
  \bibinfo{pages}{1--60}.

\bibitem[{Qin and Roberts(2020)}]{Qin:2020rad}
\bibinfo{author}{S.-X. Qin}, \bibinfo{author}{C.~D. Roberts},
  \bibinfo{title}{{Impressions of the Continuum Bound State Problem in QCD}},
  \bibinfo{journal}{Chin. Phys. Lett.}
  \bibinfo{volume}{37}~(\bibinfo{number}{12}) (\bibinfo{year}{2020})
  \bibinfo{pages}{121201}.

\bibitem[{Qin et~al.(2011)Qin, Chang, Liu, Roberts, and Wilson}]{Qin:2011dd}
\bibinfo{author}{S.-X. Qin}, \bibinfo{author}{L.~Chang}, \bibinfo{author}{Y.-X.
  Liu}, \bibinfo{author}{C.~D. Roberts}, \bibinfo{author}{D.~J. Wilson},
  \bibinfo{title}{{Interaction model for the gap equation}},
  \bibinfo{journal}{Phys. Rev. C} \bibinfo{volume}{84} (\bibinfo{year}{2011})
  \bibinfo{pages}{042202(R)}.

\bibitem[{Qin et~al.(2012)Qin, Chang, Liu, Roberts, and Wilson}]{Qin:2011xq}
\bibinfo{author}{S.-X. Qin}, \bibinfo{author}{L.~Chang}, \bibinfo{author}{Y.-x.
  Liu}, \bibinfo{author}{C.~D. Roberts}, \bibinfo{author}{D.~J. Wilson},
  \bibinfo{title}{{Investigation of rainbow-ladder truncation for excited and
  exotic mesons}}, \bibinfo{journal}{Phys. Rev. C} \bibinfo{volume}{85}
  (\bibinfo{year}{2012}) \bibinfo{pages}{035202}.

\bibitem[{Qin et~al.(2018)Qin, Roberts, and Schmidt}]{Qin:2018dqp}
\bibinfo{author}{S.-X. Qin}, \bibinfo{author}{C.~D. Roberts},
  \bibinfo{author}{S.~M. Schmidt}, \bibinfo{title}{{Poincar{\'e}-covariant
  analysis of heavy-quark baryons}}, \bibinfo{journal}{Phys. Rev. D}
  \bibinfo{volume}{97} (\bibinfo{year}{2018}) \bibinfo{pages}{114017}.

\bibitem[{Binosi et~al.(2015)Binosi, Chang, Papavassiliou, and
  Roberts}]{Binosi:2014aea}
\bibinfo{author}{D.~Binosi}, \bibinfo{author}{L.~Chang},
  \bibinfo{author}{J.~Papavassiliou}, \bibinfo{author}{C.~D. Roberts},
  \bibinfo{title}{{Bridging a gap between continuum-QCD and ab initio
  predictions of hadron observables}}, \bibinfo{journal}{Phys. Lett. B}
  \bibinfo{volume}{742} (\bibinfo{year}{2015}) \bibinfo{pages}{183--188}.

\bibitem[{Wang et~al.(2018)Wang, Qin, Roberts, and Schmidt}]{Wang:2018kto}
\bibinfo{author}{Q.-W. Wang}, \bibinfo{author}{S.-X. Qin},
  \bibinfo{author}{C.~D. Roberts}, \bibinfo{author}{S.~M. Schmidt},
  \bibinfo{title}{{Proton tensor charges from a Poincar{\'e}-covariant Faddeev
  equation}}, \bibinfo{journal}{Phys. Rev. D} \bibinfo{volume}{98}
  (\bibinfo{year}{2018}) \bibinfo{pages}{054019}.

\bibitem[{Ding et~al.(2019)Ding, Raya, Bashir, Binosi, Chang, Chen, and
  Roberts}]{Ding:2018xwy}
\bibinfo{author}{M.~Ding}, \bibinfo{author}{K.~Raya},
  \bibinfo{author}{A.~Bashir}, \bibinfo{author}{D.~Binosi},
  \bibinfo{author}{L.~Chang}, \bibinfo{author}{M.~Chen}, \bibinfo{author}{C.~D.
  Roberts}, \bibinfo{title}{{$\gamma^\ast \gamma \to \eta, \eta^\prime$
  transition form factors}}, \bibinfo{journal}{Phys. Rev. D}
  \bibinfo{volume}{99} (\bibinfo{year}{2019}) \bibinfo{pages}{014014}.

\bibitem[{Binosi et~al.(2019)Binosi, Chang, Ding, Gao, Papavassiliou, and
  Roberts}]{Binosi:2018rht}
\bibinfo{author}{D.~Binosi}, \bibinfo{author}{L.~Chang},
  \bibinfo{author}{M.~Ding}, \bibinfo{author}{F.~Gao},
  \bibinfo{author}{J.~Papavassiliou}, \bibinfo{author}{C.~D. Roberts},
  \bibinfo{title}{{Distribution Amplitudes of Heavy-Light Mesons}},
  \bibinfo{journal}{Phys. Lett. B} \bibinfo{volume}{790} (\bibinfo{year}{2019})
  \bibinfo{pages}{257--262}.

\bibitem[{Qin et~al.(2019)Qin, Roberts, and Schmidt}]{Qin:2019hgk}
\bibinfo{author}{S.-X. Qin}, \bibinfo{author}{C.~D. Roberts},
  \bibinfo{author}{S.~M. Schmidt}, \bibinfo{title}{{Spectrum of light- and
  heavy-baryons}}, \bibinfo{journal}{Few Body Syst.} \bibinfo{volume}{60}
  (\bibinfo{year}{2019}) \bibinfo{pages}{26}.

\bibitem[{Xu et~al.(2019)Xu, Binosi, Cui, Li, Roberts, Xu, and
  Zong}]{Xu:2019ilh}
\bibinfo{author}{Y.-Z. Xu}, \bibinfo{author}{D.~Binosi}, \bibinfo{author}{Z.-F.
  Cui}, \bibinfo{author}{B.-L. Li}, \bibinfo{author}{C.~D. Roberts},
  \bibinfo{author}{S.-S. Xu}, \bibinfo{author}{H.-S. Zong},
  \bibinfo{title}{{Elastic electromagnetic form factors of vector mesons}},
  \bibinfo{journal}{Phys. Rev. D} \bibinfo{volume}{100} (\bibinfo{year}{2019})
  \bibinfo{pages}{114038}.

\bibitem[{Yao et~al.(2020)Yao, Binosi, Cui, Roberts, Xu, and
  Zong}]{Yao:2020vef}
\bibinfo{author}{Z.-Q. Yao}, \bibinfo{author}{D.~Binosi},
  \bibinfo{author}{Z.-F. Cui}, \bibinfo{author}{C.~D. Roberts},
  \bibinfo{author}{S.-S. Xu}, \bibinfo{author}{H.-S. Zong},
  \bibinfo{title}{{Semileptonic decays of $D_{(s)}$ mesons}},
  \bibinfo{journal}{Phys. Rev. D} \bibinfo{volume}{102} (\bibinfo{year}{2020})
  \bibinfo{pages}{014007}.

\bibitem[{Munczek(1995)}]{Munczek:1994zz}
\bibinfo{author}{H.~J. Munczek}, \bibinfo{title}{{Dynamical chiral symmetry
  breaking, Goldstone's theorem and the consistency of the Schwinger-Dyson and
  Bethe-Salpeter Equations}}, \bibinfo{journal}{Phys. Rev. D}
  \bibinfo{volume}{52} (\bibinfo{year}{1995}) \bibinfo{pages}{4736--4740}.

\bibitem[{Bender et~al.(1996)Bender, Roberts, and von Smekal}]{Bender:1996bb}
\bibinfo{author}{A.~Bender}, \bibinfo{author}{C.~D. Roberts},
  \bibinfo{author}{L.~von Smekal}, \bibinfo{title}{{Goldstone Theorem and
  Diquark Confinement Beyond Rainbow- Ladder Approximation}},
  \bibinfo{journal}{Phys. Lett. B} \bibinfo{volume}{380} (\bibinfo{year}{1996})
  \bibinfo{pages}{7--12}.

\bibitem[{Ji and Maris(2001)}]{Ji:2001pj}
\bibinfo{author}{C.-R. Ji}, \bibinfo{author}{P.~Maris},
  \bibinfo{title}{{$K_{\ell3}$ transition form-factors}},
  \bibinfo{journal}{Phys. Rev.} \bibinfo{volume}{D64} (\bibinfo{year}{2001})
  \bibinfo{pages}{014032}.

\bibitem[{Chang et~al.(2009)Chang, Liu, Roberts, Shi, Sun, and
  Zong}]{Chang:2008ec}
\bibinfo{author}{L.~Chang}, \bibinfo{author}{Y.-X. Liu}, \bibinfo{author}{C.~D.
  Roberts}, \bibinfo{author}{Y.-M. Shi}, \bibinfo{author}{W.-M. Sun},
  \bibinfo{author}{H.-S. Zong}, \bibinfo{title}{{Chiral susceptibility and the
  scalar Ward identity}}, \bibinfo{journal}{Phys. Rev. C} \bibinfo{volume}{79}
  (\bibinfo{year}{2009}) \bibinfo{pages}{035209}.

\bibitem[{Binosi et~al.(2017)Binosi, Roberts, and
  Rodr{\'i}guez-Quintero}]{Binosi:2016xxu}
\bibinfo{author}{D.~Binosi}, \bibinfo{author}{C.~D. Roberts},
  \bibinfo{author}{J.~Rodr{\'i}guez-Quintero}, \bibinfo{title}{{Scale-setting,
  flavour dependence and chiral symmetry restoration}}, \bibinfo{journal}{Phys.
  Rev. D} \bibinfo{volume}{95} (\bibinfo{year}{2017}) \bibinfo{pages}{114009}.

\bibitem[{Gao et~al.(2018)Gao, Qin, Roberts, and
  Rodr{\'{\i}}guez-Quintero}]{Gao:2017uox}
\bibinfo{author}{F.~Gao}, \bibinfo{author}{S.-X. Qin}, \bibinfo{author}{C.~D.
  Roberts}, \bibinfo{author}{J.~Rodr{\'{\i}}guez-Quintero},
  \bibinfo{title}{{Locating the Gribov horizon}}, \bibinfo{journal}{Phys. Rev.
  D} \bibinfo{volume}{97} (\bibinfo{year}{2018}) \bibinfo{pages}{034010}.

\bibitem[{Cui et~al.(2020{\natexlab{a}})Cui, Zhang, Binosi, de~Soto, Mezrag,
  Papavassiliou, Roberts, Rodr{\'{\i}}guez-Quintero, Segovia, and
  Zafeiropoulos}]{Cui:2019dwv}
\bibinfo{author}{Z.-F. Cui}, \bibinfo{author}{J.-L. Zhang},
  \bibinfo{author}{D.~Binosi}, \bibinfo{author}{F.~de~Soto},
  \bibinfo{author}{C.~Mezrag}, \bibinfo{author}{J.~Papavassiliou},
  \bibinfo{author}{C.~D. Roberts},
  \bibinfo{author}{J.~Rodr{\'{\i}}guez-Quintero}, \bibinfo{author}{J.~Segovia},
  \bibinfo{author}{S.~Zafeiropoulos}, \bibinfo{title}{{Effective charge from
  lattice QCD}}, \bibinfo{journal}{Chin. Phys. C} \bibinfo{volume}{44}
  (\bibinfo{year}{2020}{\natexlab{a}}) \bibinfo{pages}{083102}.

\bibitem[{K\i{}z\i{}lers\"u et~al.(2021)K\i{}z\i{}lers\"u, Oliveira, Silva,
  Skullerud, and Sternbeck}]{Kizilersu:2021jen}
\bibinfo{author}{A.~K\i{}z\i{}lers\"u}, \bibinfo{author}{O.~Oliveira},
  \bibinfo{author}{P.~J. Silva}, \bibinfo{author}{J.-I. Skullerud},
  \bibinfo{author}{A.~Sternbeck}, \bibinfo{title}{{Quark-gluon vertex from Nf=2
  lattice QCD -- arXiv:2103.02945 [hep-lat]}} .

\bibitem[{Roberts(2020)}]{Roberts:2020hiw}
\bibinfo{author}{C.~D. Roberts}, \bibinfo{title}{{Empirical Consequences of
  Emergent Mass}}, \bibinfo{journal}{Symmetry} \bibinfo{volume}{12}
  (\bibinfo{year}{2020}) \bibinfo{pages}{1468}.

\bibitem[{Roberts et~al.(2021)Roberts, Richards, Horn, and
  Chang}]{Roberts:2021nhw}
\bibinfo{author}{C.~D. Roberts}, \bibinfo{author}{D.~G. Richards},
  \bibinfo{author}{T.~Horn}, \bibinfo{author}{L.~Chang},
  \bibinfo{title}{{\emph{Insights into the Emergence of Mass from Studies of
  Pion and Kaon Structure} -- arXiv:2102.01765 [hep-ph]}},
  \bibinfo{journal}{Prog. Part. Nucl. Phys.} \bibinfo{volume}{\emph{in press}}.

\bibitem[{Davies et~al.(2010)}]{Davies:2010ip}
\bibinfo{author}{C.~T.~H. Davies}, et~al., \bibinfo{title}{{Update: Precision
  $D_s$ decay constant from full lattice QCD using very fine lattices}},
  \bibinfo{journal}{Phys. Rev. D} \bibinfo{volume}{82} (\bibinfo{year}{2010})
  \bibinfo{pages}{114504}.

\bibitem[{McNeile et~al.(2012)McNeile, Davies, Follana, Hornbostel, and
  Lepage}]{McNeile:2012qf}
\bibinfo{author}{C.~McNeile}, \bibinfo{author}{C.~T.~H. Davies},
  \bibinfo{author}{E.~Follana}, \bibinfo{author}{K.~Hornbostel},
  \bibinfo{author}{G.~P. Lepage}, \bibinfo{title}{{Heavy meson masses and decay
  constants from relativistic heavy quarks in full lattice QCD}},
  \bibinfo{journal}{Phys. Rev. D} \bibinfo{volume}{86} (\bibinfo{year}{2012})
  \bibinfo{pages}{074503}.

\bibitem[{Colquhoun et~al.(2015)Colquhoun, Davies, Dowdall, Kettle, Koponen,
  Lepage, and Lytle}]{Colquhoun:2015oha}
\bibinfo{author}{B.~Colquhoun}, \bibinfo{author}{C.~T.~H. Davies},
  \bibinfo{author}{R.~J. Dowdall}, \bibinfo{author}{J.~Kettle},
  \bibinfo{author}{J.~Koponen}, \bibinfo{author}{G.~P. Lepage},
  \bibinfo{author}{A.~T. Lytle}, \bibinfo{title}{{B-meson decay constants: a
  more complete picture from full lattice QCD}}, \bibinfo{journal}{Phys. Rev.
  D} \bibinfo{volume}{91} (\bibinfo{year}{2015}) \bibinfo{pages}{114509}.

\bibitem[{Maris and Roberts(1997)}]{Maris:1997tm}
\bibinfo{author}{P.~Maris}, \bibinfo{author}{C.~D. Roberts},
  \bibinfo{title}{{{$\pi$} and {$K$} meson Bethe-Salpeter amplitudes}},
  \bibinfo{journal}{Phys. Rev. C} \bibinfo{volume}{56} (\bibinfo{year}{1997})
  \bibinfo{pages}{3369--3383}.

\bibitem[{Windisch(2017)}]{Windisch:2016iud}
\bibinfo{author}{A.~Windisch}, \bibinfo{title}{{Analytic properties of the
  quark propagator from an effective infrared interaction model}},
  \bibinfo{journal}{Phys. Rev. C} \bibinfo{volume}{95} (\bibinfo{year}{2017})
  \bibinfo{pages}{045204}.

\bibitem[{Chang et~al.(2013)Chang, Cloet, Roberts, Schmidt, and
  Tandy}]{Chang:2013nia}
\bibinfo{author}{L.~Chang}, \bibinfo{author}{I.~C. Cloet},
  \bibinfo{author}{C.~D. Roberts}, \bibinfo{author}{S.~M. Schmidt},
  \bibinfo{author}{P.~C. Tandy}, \bibinfo{title}{{Pion electromagnetic form
  factor at spacelike momenta}}, \bibinfo{journal}{Phys. Rev. Lett.}
  \bibinfo{volume}{111} (\bibinfo{year}{2013}) \bibinfo{pages}{141802}.

\bibitem[{Nakanishi(1969)}]{Nakanishi:1969ph}
\bibinfo{author}{N.~Nakanishi}, \bibinfo{title}{{A General survey of the theory
  of the Bethe-Salpeter equation}}, \bibinfo{journal}{Prog. Theor. Phys.
  Suppl.} \bibinfo{volume}{43} (\bibinfo{year}{1969}) \bibinfo{pages}{1--81}.

\bibitem[{Chen et~al.(2019)Chen, Lu, Binosi, Roberts, Rodr\'\i{}guez-Quintero,
  and Segovia}]{Chen:2018nsg}
\bibinfo{author}{C.~Chen}, \bibinfo{author}{Y.~Lu},
  \bibinfo{author}{D.~Binosi}, \bibinfo{author}{C.~D. Roberts},
  \bibinfo{author}{J.~Rodr\'\i{}guez-Quintero}, \bibinfo{author}{J.~Segovia},
  \bibinfo{title}{{Nucleon-to-Roper electromagnetic transition form factors at
  large $Q^2$}}, \bibinfo{journal}{Phys. Rev. D} \bibinfo{volume}{99}
  (\bibinfo{year}{2019}) \bibinfo{pages}{034013}.

\bibitem[{Binosi and Tripolt(2020)}]{Binosi:2019ecz}
\bibinfo{author}{D.~Binosi}, \bibinfo{author}{R.-A. Tripolt},
  \bibinfo{title}{{Spectral functions of confined particles}},
  \bibinfo{journal}{Phys. Lett. B} \bibinfo{volume}{801} (\bibinfo{year}{2020})
  \bibinfo{pages}{135171}.

\bibitem[{Ding et~al.(2020{\natexlab{a}})Ding, Raya, Binosi, Chang, Roberts,
  and Schmidt}]{Ding:2019lwe}
\bibinfo{author}{M.~Ding}, \bibinfo{author}{K.~Raya},
  \bibinfo{author}{D.~Binosi}, \bibinfo{author}{L.~Chang},
  \bibinfo{author}{C.~D. Roberts}, \bibinfo{author}{S.~M. Schmidt},
  \bibinfo{title}{{Symmetry, symmetry breaking, and pion parton
  distributions}}, \bibinfo{journal}{Phys. Rev. D}
  \bibinfo{volume}{101}~(\bibinfo{number}{5})
  (\bibinfo{year}{2020}{\natexlab{a}}) \bibinfo{pages}{054014}.

\bibitem[{Ding et~al.(2020{\natexlab{b}})Ding, Raya, Binosi, Chang, Roberts,
  and Schmidt}]{Ding:2019qlr}
\bibinfo{author}{M.~Ding}, \bibinfo{author}{K.~Raya},
  \bibinfo{author}{D.~Binosi}, \bibinfo{author}{L.~Chang},
  \bibinfo{author}{C.~D. Roberts}, \bibinfo{author}{S.~M. Schmidt},
  \bibinfo{title}{{Drawing insights from pion parton distributions}},
  \bibinfo{journal}{Chin. Phys. C (Lett.)} \bibinfo{volume}{44}
  (\bibinfo{year}{2020}{\natexlab{b}}) \bibinfo{pages}{031002}.

\bibitem[{Souza et~al.(2020)Souza, Narciso~Ferreira, Aguilar, Papavassiliou,
  Roberts, and Xu}]{Souza:2019ylx}
\bibinfo{author}{E.~V. Souza}, \bibinfo{author}{M.~Narciso~Ferreira},
  \bibinfo{author}{A.~C. Aguilar}, \bibinfo{author}{J.~Papavassiliou},
  \bibinfo{author}{C.~D. Roberts}, \bibinfo{author}{S.-S. Xu},
  \bibinfo{title}{{Pseudoscalar glueball mass: a window on three-gluon
  interactions}}, \bibinfo{journal}{Eur. Phys. J. A (Lett.)}
  \bibinfo{volume}{56} (\bibinfo{year}{2020}) \bibinfo{pages}{25}.

\bibitem[{Cui et~al.(2020{\natexlab{b}})Cui, Chen, Binosi, de~Soto, Roberts,
  Rodr{\'{\i}}guez-Quintero, Schmidt, and Segovia}]{Cui:2020rmu}
\bibinfo{author}{Z.-F. Cui}, \bibinfo{author}{C.~Chen},
  \bibinfo{author}{D.~Binosi}, \bibinfo{author}{F.~de~Soto},
  \bibinfo{author}{C.~D. Roberts},
  \bibinfo{author}{J.~Rodr{\'{\i}}guez-Quintero}, \bibinfo{author}{S.~M.
  Schmidt}, \bibinfo{author}{J.~Segovia}, \bibinfo{title}{{Nucleon elastic form
  factors at accessible large spacelike momenta}}, \bibinfo{journal}{Phys. Rev.
  D} \bibinfo{volume}{102} (\bibinfo{year}{2020}{\natexlab{b}})
  \bibinfo{pages}{014043}.

\bibitem[{Huber et~al.(2020)Huber, Fischer, and Sanchis-Alepuz}]{Huber:2020ngt}
\bibinfo{author}{M.~Q. Huber}, \bibinfo{author}{C.~S. Fischer},
  \bibinfo{author}{H.~Sanchis-Alepuz}, \bibinfo{title}{{Spectrum of scalar and
  pseudoscalar glueballs from functional methods}}, \bibinfo{journal}{Eur.
  Phys. J. C} \bibinfo{volume}{80}~(\bibinfo{number}{11})
  (\bibinfo{year}{2020}) \bibinfo{pages}{1077}.

\bibitem[{Cui et~al.(2021)Cui, Binosi, Roberts, and Schmidt}]{Cui:2021vgm}
\bibinfo{author}{Z.-F. Cui}, \bibinfo{author}{D.~Binosi},
  \bibinfo{author}{C.~D. Roberts}, \bibinfo{author}{S.~M. Schmidt},
  \bibinfo{title}{{\emph{Fresh extraction of the proton charge radius from
  electron scattering} -- arXiv:2102.01180 [hep-ph]}} .

\bibitem[{Schlessinger and Schwartz(1966)}]{Schlessinger:1966zz}
\bibinfo{author}{L.~Schlessinger}, \bibinfo{author}{C.~Schwartz},
  \bibinfo{title}{{Analyticity as a Useful Computation Tool}},
  \bibinfo{journal}{Phys. Rev. Lett.} \bibinfo{volume}{16}
  (\bibinfo{year}{1966}) \bibinfo{pages}{1173--1174}.

\bibitem[{Schlessinger(1968)}]{PhysRev.167.1411}
\bibinfo{author}{L.~Schlessinger}, \bibinfo{title}{Use of Analyticity in the
  Calculation of Nonrelativistic Scattering Amplitudes},
  \bibinfo{journal}{Phys. Rev.} \bibinfo{volume}{167} (\bibinfo{year}{1968})
  \bibinfo{pages}{1411--1423}.

\bibitem[{Mathur et~al.(2018)Mathur, Padmanath, and Mondal}]{Mathur:2018epb}
\bibinfo{author}{N.~Mathur}, \bibinfo{author}{M.~Padmanath},
  \bibinfo{author}{S.~Mondal}, \bibinfo{title}{{Precise predictions of
  charmed-bottom hadrons from lattice QCD}}, \bibinfo{journal}{Phys. Rev.
  Lett.} \bibinfo{volume}{121}~(\bibinfo{number}{20}) (\bibinfo{year}{2018})
  \bibinfo{pages}{202002}.

\bibitem[{Hu et~al.(2020{\natexlab{b}})Hu, Jin, and Xiao}]{Hu:2019bdf}
\bibinfo{author}{X.-Q. Hu}, \bibinfo{author}{S.-P. Jin}, \bibinfo{author}{Z.-J.
  Xiao}, \bibinfo{title}{{Semileptonic decays $B/B_s \to (D^{(*)},D_s^{(*)}) l
  \nu_l$ in the PQCD approach with the lattice QCD input}},
  \bibinfo{journal}{Chin. Phys. C} \bibinfo{volume}{44}~(\bibinfo{number}{5})
  (\bibinfo{year}{2020}{\natexlab{b}}) \bibinfo{pages}{053102}.

\bibitem[{Gambino et~al.(2020)}]{Gambino:2020jvv}
\bibinfo{author}{P.~Gambino}, et~al., \bibinfo{title}{{Challenges in
  semileptonic $B$ decays}}, \bibinfo{journal}{Eur. Phys. J. C}
  \bibinfo{volume}{80}~(\bibinfo{number}{10}) (\bibinfo{year}{2020})
  \bibinfo{pages}{966}.

\end{thebibliography}

\end{document}